\documentclass[sigconf]{acmart}

\AtBeginDocument{%
  \providecommand\BibTeX{{%
    \normalfont B\kern-0.5em{\scshape i\kern-0.25em b}\kern-0.8em\TeX}}}
\renewcommand\footnotetextcopyrightpermission[1]{}
\usepackage{multirow}
\usepackage{algorithm}
\usepackage{algorithmic}
\usepackage{balance}
\usepackage[T1]{fontenc}
\begin{document}

\title{$\lambda$-domain  VVC Rate Control  Based on Game Theory}

\author{Jielian Lin}
\affiliation{%
	\institution{Fujian Key Lab for Intelligent Processing and Wireless Transmission of Media Information, Fuzhou University}
	\city{Fuzhou}
	\country{China}}
\author{Aiping Huang}
\affiliation{%
	\institution{Fujian Key Lab for Intelligent Processing and Wireless Transmission of Media Information, Fuzhou University}
	\city{Fuzhou}
	\country{China}}

\author{Keke Zhang}
\affiliation{%
	\institution{Fujian Key Lab for Intelligent Processing and Wireless Transmission of Media Information, Fuzhou University}
	\city{Fuzhou}
	\country{China}}

\author{Xu Wang}
\affiliation{%
	\institution{College of Computer Science and Software Engineering, Shenzhen University}
	\city{Shenzhen}
	\country{China}}

\author{Tiesong Zhao}

\affiliation{%
	\institution{Fujian Key Lab for Intelligent Processing and Wireless Transmission of Media Information, Fuzhou University}
	\city{Fuzhou}
	\country{China}
}

\email{t.zhao@fzu.edu.cn}


\begin{abstract}
 Versatile Video Coding (VVC)  has set a new milestone in high-efficiency video coding. In the standard encoder, the $\lambda$-domain rate control is incorporated for its high accuracy and good Rate-Distortion (RD) performance. In this paper, we formulate this task as a Nash equilibrium problem that effectively bargains between multiple agents, {\it i.e.}, Coding Tree Units (CTUs) in the  frame. After that, we calculate the optimal $\lambda$ value with a two-step strategy: a Newton method to iteratively obtain an intermediate variable, and a solution of Nash equilibrium to obtain the optimal $\lambda$. Finally, we propose an effective CTU-level rate allocation with the optimal $\lambda$ value. To the best of our knowledge, we are the first to combine game theory with $\lambda$-domain rate control. Experimental results with Common Test Conditions (CTC) demonstrate the efficiency of the proposed method, which outperforms the state-of-the-art CTU-level rate allocation algorithms.
\end{abstract}




\maketitle
\section{Introduction}
To date, different types of videos have been widely used in broadcasting, teaching, entertainment, and so on. With the explosive growth of   video consumption, efficient video compression and delivery have become urgent problems. In the past decades, video coding standards (\textit{e.g.}, H.264/AVC \cite{introduceH264}, High Efficiency Video Coding (H.265/HEVC) \cite{introduceH265}, and Versatile Video Coding (VVC) \cite{introduceVVC}) have  been developed to solve these problems in different periods.   The up-to-date standard VVC,  published in 2020 by the Joint Video Experts Team (JVET), has enhanced the coding performance of H.265/HEVC by reducing its bit rate by 50\% at the same coding quality. However,  as there is limited bandwidth and storage to transmit video data, Rate Control (RC) \cite{RC-Li20,TCSVT21-Liu,Rc-mao21}  is still one of the major research areas in the VVC standard.

\begin{figure}[t]
	\centering
	\includegraphics [width=0.5\textwidth] {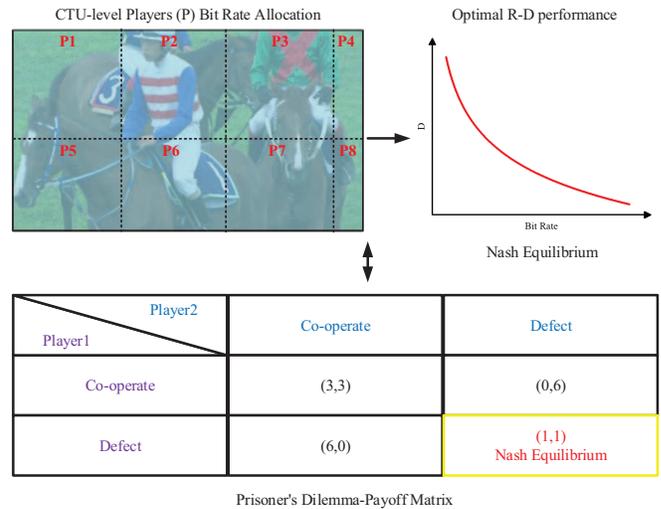}	
	\caption{CTU-level  Cooperative Bargaining Game.}
	\label{framework}
\end{figure}

In video coding, the primary problem of RC is to maximize the encoded video's perceptual quality under a constrained bit rate. In the past decades,  many approaches have been proposed to enhance the Rate Distortion (RD) performance by adjusting the Quantization Parameter (Qp), bit rate, or Lagrange multiplier parameters. For H.264/AVC and H.265/HEVC, many methods (\textit{e.g.}, the $\lambda$-domain RD model \cite{slidewindowsLi2014,VideoRD-Li16,VideoRD-Li18}, deep-learning-based quantization parameters prediction \cite{Videomachinelearning-Zhou21,Videomachinelearning-ho21},   perception-quality-based \cite{VideoSSIM-Zhao15,VideoJND-Zhou20,TBC21-Li} bit rate allocation)  have been proposed to minimize compression distortion under a constrained bit rate. While in the current VVC reference software VTM13.0 \cite{VVC}, the $\lambda$-domain RD model with the outstanding RD performance was set as the default RC scheme. To date, the RC methods \cite{Rc-mao21,RC-Wang20} for VVC are 
more inclined to precisely characterize the relationship between R and D. However, the bit allocation strategy based on the $\lambda$-domain RD model has not been explored for all uncoded CTUs in an updated way. Although there is a sliding window to smooth the bit rate of CTUs in the window,  it lacks sufficient theoretical support to balance  bit allocation of all uncoded CTUs to achieve globally optimized coding performance. Therefore, it is still necessary to develop a new method to enhance the RD performance in video coding.

Game theory is an efficient strategy to balance  fairness and efficiency in the case of opaque information for the limited resource allocation problem (\textit{e.g.}, the one-pass RC problem in HEVC). Some studies have implied that game theory can perform well in the CTU-based \cite{GTAhmad2006,GTGAO2017}, band-based \cite{GTACMMM-Zhao21},  and frame-based \cite{GTwang2013} bit rate allocations in H.264 or HEVC. In these methods, the bit rate allocation for different levels of coding unit was defined as a Nash equilibrium problem. Although these methods have presented persuasive performance, all of them were designed for  the RC in H.264 or HEVC. In addition, the CTU-level bit rate allocation based on the outstanding $\lambda$-domain RD model  has not been explored  and it still is  a cooperative bargaining game problem, as shown in Fig. \ref{framework}. Therefore,  the  $\lambda$-domain RC based on game theory is encouraged and has the potential to enhance the  RD performance in VVC.

Based on the above investigation, we formulate the CTU-level RC in VVC as Nash equilibrium problem and attempt to address it by considering the popular $\lambda$- domain RD model. In summary, the major contributions of this paper are presented as follows:
\begin{itemize}
	\item 	 We are the first to address the $\lambda$-domain RC problem with game theory.  We formulate the one-pass CTU-level RC problem  with the popular $\lambda$-domain RD model.  
	\item  We obtain the optimal $\lambda$ value with a two-step strategy: a Newton method to iteratively obtain an intermediate variable, and a solution of Nash equilibrium to obtain the optimal $\lambda$. 
	\item We propose an overall RC framework with the optimal $\lambda$ and implement in the newest video coding standard H.266/VVC. Experimental results demonstrate that the proposed method outperforms  benchmarks. 
\end{itemize}

\section{Related work}
\label{relatedwork}

\subsection{Rate Control for VVC}
Nowadays, to enhance the coding performance of VVC, Mao \textit{et al.} \cite{Rc-mao21} exploited RD characteristics for sequences by deriving new R-Q and D-Q models for frame-level RC. A dependency factor models the inter-frame dependency characteristics. Then, they proposed an adaptive bit allocation method based on the R-Q model, D-Q model, and the dependency factor. In \cite{RC-Li20}, Li \textit{et al.}  proposed an RD parameter updating strategy and explored the quality dependency between frames. In \cite{framelevelHyun}, a frame-level constant bit-rate (CBR) control method using Recursive Bayesian Estimation (RBE) was proposed. The RBE with alternating prediction and updated steps could  estimate  bit rates and  allocate target bit rates based on the distortions changes of the previously coded frames. Wang \textit{et al.} \cite{RC-Wang20} shifted the traditional bit rate-centered paradigm to a quality-centered paradigm. In addition, they developed the relationship between   Structural Similarity Index Metric (SSIM) and Qp for quality control. Liu \textit{et al.} \cite{TCSVT21-Liu} proposed a multi-objective optimization of quality-based CTU level RC method. Their method converts the multi-objective optimization problem into a single-objective problem to minimize average distortion and quality fluctuation for the constrained bit rate. Then, a two-stage method based on the D-$\lambda$ model was proposed to obtain the optimal solution. In this method, the obtained solution allocates the bit rate to all CTUs once for all.   

\subsection{$\lambda$-domain $RD$ model}
In \cite{slidewindowsLi2014}, the $\lambda$-domain RC was firstly proposed and integrated into HEVC. It provided a R-$\lambda$ model that can better characterize the relationship between R and D than  traditional models. Then, to achieve the optimal RD performance, Li \textit{et al.} \cite{VideoRD-Li18} further proposed a D-$\lambda$ model to characterize the RD behavior of video content. Based on the $\lambda$-domain RD model,  many approaches were proposed. Zhou \textit{et al.} \cite{R-LAMBDA-ZHOU2020} proposed a High Dynamic Range (HDR)-visual difference predictor-2-based RD model to improve video coding performance by considering the HDR characteristics. In \cite{R-LAMBDA-Li2020}, Li \textit{et al.} designed an optimal CTU level weight for the equi-rectangle projection format of the 360-degree video and proposed a weighted CTU level bit allocation method. Recently, Li \textit{et al.} \cite{TBC21-Li} proposed an SSIM-based optimal bit allocation and an SSIM-based RD optimization (SOSR)  to solve the inconsistency between optimal bit allocation and RD optimization results in non-optimal SSIM-based coding. In \cite{R-lambda-guotbc20}, by considering the relationship between weighted Lagrange multiplier and temporal dependency, Guo \textit{et al.} proposed a CTU level RD model. They also developed a formulation by combining inter-block dependency and RD characteristics.

\subsection{Game Theory for Video Coding}
Game theory can solve the problem of how to achieve both fairness and efficiency in the case of opaque information. Based on this theory,  Ahmad \textit{et al.} \cite{GTAhmad2006} first defined bit rate allocation between macroblocks in a frame as a cooperative game.  In \cite{gt-wangtip14},  Generalized Nash Bargaining Solution (GNBS) was developed to solve the inner-layer bit allocation of H.264.  Wang \textit{et al.} \cite{GTwang2013} modeled frame-level bit rate allocation in different temporal layers as a game theory problem. They also analyzed the relationship between different temporal layers and employed it to define the utility function for Nash equilibrium. Based on the statistical characteristics of panoramic pictures, Zhao \textit{et al.} \cite{GTACMMM-Zhao21}  introduced the game theory to the RC for 360-degree video coding and allocated the bit rate band by band.  Gao \textit{et al.} \cite{GTGAO2017} adopted support vector machine-based multi-classification scheme to determine RD model for all CTUs. They also designed the bit rate allocation scheme based on the mixed RD model-based cooperative bargaining game theory.  The performance of these works has verified the efficiency of game theory for the one-pass RC problem.  

To date, the game theory has been adopted in the RC at  different levels of coding units, such as  CTU \cite{GTAhmad2006,GTGAO2017},  band \cite{GTACMMM-Zhao21}, and frame \cite{GTwang2013,gt-wangtip14}. In such a bargaining game  of RC,  each player is expected to obtain a fair share of total bit rate whilst minimizing  the overall distortion. In other words, each player needs to reach a beneficial agreement. It is worthy to notice that if this agreement has been reached, every player has its utility function $u_{p}$, where $p$ is the player's index. However, if such an agreement is impossible, a minimum utilities $u_{p}^0$ for every player. These minimum utilities  are denoted as the minimum utility pair $u^{0}=(u_{1}^0,u_{2}^0,\cdots,u_{p}^0)$. As there is a feasible utility set $U$, which is all the possible utility pairs, for the bargaining game, the Pareto optimality solution $u^{*}=(u_{1}^{*},u_{2}^{*},\cdots,u_{p}^{*})$ can be obtained by the Nash Bargaining Solution (NBS) \cite{GTNash50} in these works. $u^{*}$ can be  represented as:
\begin{equation}
\label{eq:NBS}
\begin{aligned}
u^{*}&=H\left ( U,u^{0} \right )=arg\max_{\left \langle U,u^{0} \right \rangle }  \prod_{p=1}^{P} \left ( u_{p}-u_{p}^0 \right ) ,
\end{aligned}
\end{equation}
where $P$ is the number of players in the barganing game.

\section{proposed Method}
\label{method}
\subsection{Problem Formulation}

Most of the RC methods for VVC optimize the RD performance by formulating a new model. However,  the default  $\lambda$-domain RD model  has  not been explored to balance the bit rate for all uncoded CTUs until now. Since some CTUs may consume too many bit rates, the remaining bits cannot obtain the minimum coding quality for uncoded CTUs. Then the overall coding performance cannot reach the optimal performance. Therefore, a bit rate allocation scheme requires to allocate a fair bit rate for all uncoded CTUs. To our best knowledge, the solution of the $\lambda$-domain RD model based on game theory for the one-pass RC has not been exploited and analyzed. In this paper,  the $\lambda$-domain RC scheme based on game theory is modeled and solved. 

The  RC problem is formulated as to minimize the total distortion, subject to a constraint on the target bit rate of compressed videos. It can be expressed as: 
\begin{equation}
\min \sum\limits_{j = 1}^{N_i} { M_{i,j}d_{i,j}}, s.t. \sum\limits_{j = 1}^{N_i} {M_{i,j}  r_{i,j}}\le  R_i,
\end{equation}
where $ M_{i,j}$ is the  number of pixels in the $j$-th CTU of the $i$-th frame.  $d_{i,j}$ is the mean distortion in the same CTU. $N_i$ is the number of uncoded CTUs in this frame. $r_{i,j}$ is the Bits Per Pixel (bpp) of this CTU and $R_i$ is the target bit rate of the $i$-th frame. 

To date, the defalut RD model of VVC regulates the relationship between $r_{i,j}$ and $d_{i,j}$ as an exponential function \cite{R-D-2012Li}, which is expressed as:
\begin{equation}
\label{eq:D}
d_{i,j}\left (r_{i,j}   \right ) =k_{i,j}r_{i,j} ^{ -c_{i,j}},
\end{equation}
where $k_{i,j}$ and $c_{i,j}$ are the model parameters  related to the contents of the video. Then, $\lambda_{i,j}$ could be obtained by:
\begin{equation}
\label{eq:lambd}
\lambda _{i,j}=-\frac{\partial d_{i,j}}{\partial r_{i,j} } =c_{i,j}k_{i,j}r_{i,j} ^{ -c_{i,j}-1}.
\end{equation}
Therefore,
\begin{equation}
\label{eq:R}
r_{i,j}=\left (\frac{\lambda_{i,j} }{c_{i,j}k_{i,j}}   \right ) ^{-\frac{1}{c_{i,j}+1} }.
\end{equation}
By substituting Eq. (\ref{eq:R}),  we can rewrite $ d_{i,j}$  as:
\begin{equation}
\label{eq:R-D}
d_{i,j}=\left ( \frac{\lambda_{i,j} k_{i,j}^{\frac{1}{c_{i,j}} } }{c_{i,j}}  \right )^{\frac{c_{i,j}}{c_{i,j}+1}  }.
\end{equation}

\begin{table}[]
	
	\caption{Key Symbol Table}
	\centering
	\label{tab:Symbol_Table}
	\scalebox{0.95}{
		\setlength{\tabcolsep}{0.9mm}{
			\begin{tabular}{|c|l|}
				\hline
				\textbf{Symbol}     & \multicolumn{1}{c|}{\textbf{Explanation}}                                                                                   \\ \hline
				$d_{i,j},r_{i,j}$   & \begin{tabular}[c]{@{}l@{}}the distortion and  bpp of the $j$-th CTU in the $i$-th  frame\end{tabular}                    \\ \hline
				$R_i$               & the target bit rate of the $i$-th frame                                                                                     \\ \hline
				$N_i$                   & the number of uncoded CTUs in the $i$-th frame                                \\ \hline
				$M_{i,j}$           & \begin{tabular}[c]{@{}l@{}}the  number of pixels of the $j$-th CTU in the  $i$-th  frame\end{tabular}                     \\ \hline
				$c_{i,j},k_{i,j}$   & \begin{tabular}[c]{@{}l@{}} the  RD model parameters in $\lambda$-domain \end{tabular}                                                                 \\ \hline
				$\lambda_{i,j}$     & the $\lambda$ value of the $j$-th CTU in the $i$-th frame                                                                   \\ \hline
				$u_{i,j},u_{i,j}^0$ & \begin{tabular}[c]{@{}l@{}}the utility  and minimal utility  of the $j$-th CTU in \\the $i$-th frame\end{tabular} \\ \hline
				$\eta$,$\eta^*$     & \begin{tabular}[c]{@{}l@{}}an intermediate variable and its optimal value to \\derive $\lambda_{i,j}$ \end{tabular}          \\ \hline
								$S_{i,j}$           & the scale factor of                                                                                                         \\ \hline
								$\varsigma$            & \begin{tabular}[c]{@{}l@{}}the ratio between the guaranteed minimum utility \\ bit rate and the  target bit rate\end{tabular}                \\ \hline
			\end{tabular}}}
		\end{table}
		
According to this RD relationship, a lower  distortion leads to a higher bit rate cost, and vice versa.  Due to the limited $R_i$,  the requirements of all CTUs cannot be satisfied to the maximum.  Therefore,    the bit rate balance is required to get the optimal distortion $\min { \sum_{j = 1}^{N_i}}  { d_{i,j}} $. This balancing procedure is a typical Nash equilibrium problem of game theory.   The objective of each player is to maximize its positive utility.  Then, the utility function of the $j$-th CTU in the $i$-th frame $u_{i,j}$ could be expressed as:
\begin{equation}
\label{eq:utility}
u_{i,j} = \frac{1}{  d_{i,j}},
\end{equation}
By substituting Eq. (\ref{eq:R-D}) into Eq. (\ref{eq:utility}), we update the utility function $u_{i,j}$ as:
\begin{equation}
\label{eq:utility2}
u_{i,j} = \left ( \frac{\lambda_{i,j} k_{i,j}^{\frac{1}{c_{i,j}} } }{c_{i,j}}  \right )^{-\frac{c_{i,j}}{c_{i,j}+1}  }.
\end{equation}
Furthermore, a minimal utility $u_{i,j}^0$ is defined to avoid severe quality degradation at each CTU. For ease of derivation, we estimate the minimal utility with information of previously coded CTUs. Therefore, it is irrelevant to any encoding parameters and outputs at $j$, \textit{i.e.},
\begin{equation}
\label{eq:uij0constraint}
\frac{\partial u_{i,j}^0}{\partial \lambda_{i,j}} = \frac{\partial u_{i,j}^0}{\partial d_{i,j}}  = 0.
\end{equation}
With the utility $u_{i,j}$ and minimal utility $u_{i,j}^0$, the objective of CTU bit rate allocation is formulated by game theory:
\begin{equation}
\label{eq:aim}
\max \sum\limits_{j = 1}^{N_i } {\ln \left( {u_{i,j} \left( {\lambda_{i,j} } \right) - u_{i,j}^0 } \right)} ,s.t.\left\{ {\begin{array}{*{20}c}
	{\sum\limits_{j = 1}^{N_i} {M_{i,j}  r_{i,j} }  \le R_i }  \\
	{r_{i,j}  > r_{i,j}^0 ,\forall j}  \\
	\end{array}} \right.,
\end{equation}
where $r_{i,j}^0$ is the required bpp for minimual utiltity $u_{i,j}^0$. For ease of derivation, the multiplication function Eq. (\ref{eq:NBS}) is changed to summation by the $\ln$($\cdot$) function. The key symbols  used for bit allocation  are summarized in Table \ref{tab:Symbol_Table}.
\subsection{Solution for the Optimized Lagrange Multiplier}

\begin{figure}[t]
	\centering
	\includegraphics [width=0.5\textwidth] {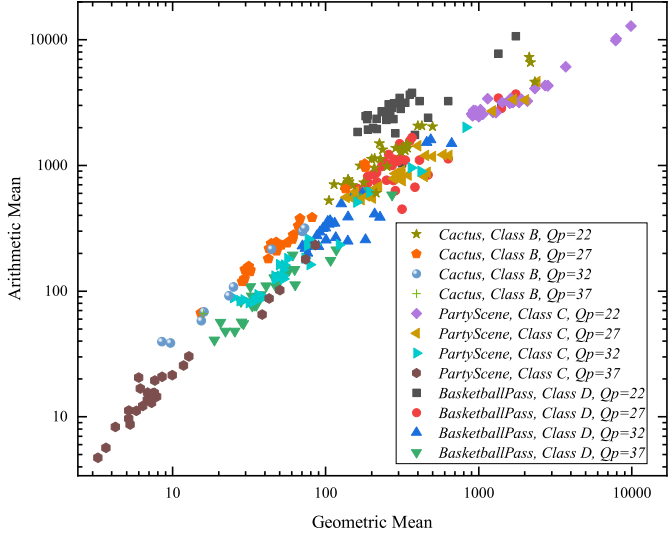}	
	\caption{The relationship between arithmetic mean and geometric mean of CTU-level bitrates.}
	\label{actgeomeanandmean}
\end{figure}

In this paper, the constraint
${\textstyle \sum_{j = 1}^{N_i}}  {M_{i,j} \cdot r_{i,j} }  \le R_i$
is relaxed to $ {\textstyle \prod_{j=1}^{N_i}} r_{i,j}\le \left ( R_{i} /(M_{i,j}\cdot N_i)   \right ) ^{N_i}  $. 
The reasons to utilize this approximation is threefold. First, the geometric and arithmetic average values are close for the allocated bit rate. As shown in Fig. \ref{actgeomeanandmean}, we summarize that the geometric mean and the arithmetic mean have similar statistical significance.  It can be readily seen that in most cases, the two values are very close to each other. Second, this approximation benefits our following derivation to obtain an optimal $\lambda_{i,j}$. Third, this approximation has been widely adopted in similar tasks with promising coding results ({\it e.g.},  \cite{tip-wang16}).
With the utility $u_{i,j}$ and minimal utility $u_{i,j}^0$, the objective of CTU bit rate allocation is updated as:
\begin{equation}
\label{eq:lm}
\begin{aligned}
F &=\sum_{j = 1}^{N_i}\ln{\left(u_{i,j}\left(\lambda_{i,j}\right)-u_{i,j}^0\right)}+\sum_{j=1}^{N_i}{\theta_{i,j}\left(r_{i,j}-r_{i,j}^0\right)}\\
&+\eta \left(\sum_{j=1}^{N_i} \ln\left ( \left (\frac{\lambda_{j} }{c_{i,j}k_{i,j}}  \right )^{-\frac{1}{c_{i,j}+1}}\right )-N_i\ln\left ( \frac{R_{i} }{M_{i,j}\cdot N_i}  \right )    \right).
\end{aligned}
\end{equation}
It can be maximized at Karush-Kuhn-Tucker (KKT)  conditions,
\begin{equation}
\label{eq:kkt}
\begin{aligned}
\left\{ {\begin{array}{ll}
	\vspace{4pt}
	{\frac{{\partial F}}{{\partial \lambda_{i,j} }} = 0,\forall j}  \\
	\vspace{4pt}
	{\theta _{i,j} \left( {r_{i,j}  - r_{i,j}^0 } \right) = 0,\forall j}   \\
	\eta \left(\sum_{j=1}^{N_i} \ln\left ( \left (\frac{\lambda_{i,j} }{c_{i,j}k_{i,j}}  \right )^{-\frac{1}{c_{i,j}+1}}\right )-N_i\ln\left ( \frac{R_{i} }{M_{i,j}\cdot N_i}  \right )    \right) = 0   \
	\end{array}} \right .. 
\end{aligned}
\end{equation}
Because $r_{i,j}-r_{i,j}^0 > 0$,  $\theta _{i,j}$ should be $0$ to ensure $\theta _{i,j} \left( {r_{i,j}  - r_{i,j}^0 } \right) = 0$. Meanwhile,  $\frac{{\partial u_{i,g}}}{\partial \lambda_{i,j} }=0, \forall g < j$. Substituting all these conclusions in $\frac{{\partial F}}{{\partial \lambda_{i,j} }}$, we get:
\begin{equation}
\label{eq:dev5}
\begin{aligned}
&\frac{{\partial F}}{{\partial \lambda_{i,j} }}  = \frac{{\partial \ln \left( {u_{i,j} \left( {\lambda_{i,j} } \right) - u_{i,j}^0 } \right)}}{{\partial \lambda _{i,j} }} -   \frac{\eta }{(c_{i,j}+1) \lambda_{i,j}}= 0.
\end{aligned}
\end{equation}
Therefore,
\begin{equation}
\label{eq:lambda}
\begin{aligned}
\lambda_{i,j} =c_{i,j}k_{i,j} ^{-\frac{1}{c_{i,j}} } \left (\frac{c_{i,j} +\eta}{\eta u_{i,j}^0} \right )   ^{\frac{c_{i,j}+1}{c_{i,j}} } , 	
\end{aligned}
\end{equation}
where $\eta \ne 0$ is known from Eq. (\ref{eq:kkt}). Then the condition is updated as:
\begin{equation}
\label{eq:condition2}
\begin{aligned}
\sum_{j=1}^{N_i} \ln\left (\left (\frac{\lambda_{i,j} }{c_{i,j}k_{i,j}}  \right )^{-\frac{1}{c_{i,j}+1}}\right )-N_i\ln\left ( \frac{R_{i,j} }{M_{i,j}\cdot N_i}  \right ) =0.\\
\end{aligned}
\end{equation}
By substituting Eq. (\ref{eq:lambda}) to Eq. (\ref{eq:condition2}), we get:
\begin{equation}
\label{eq:dev6}
\begin{aligned}	
\sum_{j=1}^{N_i}\frac{1}{c_{i,j}}\ln  \left ( \frac{c_{i,j}}{\eta }+1  \right )  &-\sum_{j=1}^{N_i}\frac{1}{c_{i,j}}\ln \left ( u_{i,j}^0k_{ij}  \right ) +N_i\ln\left ( \frac{R_{i} }{M_{i,j}  \cdot N_i}  \right )= 0.\\
\end{aligned}
\end{equation}	
Since the parameters $c_{i,j}$ of  uncoded CTUs are not constant, the colsed-form solution of Eq. (\ref{eq:dev6}) can not be got. To obtain the accurate solution of Eq. (\ref{eq:dev6}), the Newton method is applied. The detailed solution is summarized in Algorithm \ref{alg1}. Eq. (\ref{eq:dev6}) is divided into two parts: $\xi$ and $f(\eta)$.  $\xi$ is represented as:
\begin{equation}
\label{eq:rhoformula}
\begin{aligned}
\xi & = \sum_{j  = 1}^{N_i}\frac{1}{c_{i,j}}\ln \left ( u_{i,j}^0k_{ij}  \right ) -N_i\ln\left ( \frac{R_{i} }{ M_{i,j} \cdot N_i}  \right )  ,\\
\end{aligned}
\end{equation}
$f(\eta)$ is defined as:
\begin{equation}
\label{eq:etaformula}
\begin{aligned}
f(\eta) & = \sum_{j  = 1}^{N_i}\frac{1}{c_{i,j}}\ln  \left ( \frac{c_{i,j}}{\eta }+1  \right ). \\
\end{aligned}
\end{equation}
Then, the solution $\eta^*$ of Eq. (\ref{eq:dev6}) can be obtained by Algorithm \ref{alg1}, which can be considered as an intermediate variable. After that, Submitting $\eta^*$ into Eq. (\ref{eq:lambda}), we  can  obtain the optimal $\lambda_{i,j}$ as:
\begin{equation}
\label{eq:lambda2}
\begin{aligned}
\lambda_{i,j} =c_{i,j}k_{i,j} ^{-\frac{1}{c_{i,j}} } \left (\frac{c_{i,j} +\eta^*}{\eta^* u_{i,j}^0} \right )   ^{\frac{c_{i,j}+1}{c_{i,j}} } . 	
\end{aligned}
\end{equation}
\begin{algorithm}[t]
	\caption{Solution for Eq. (\ref{eq:dev6})}
	\begin{algorithmic}
		\STATE {Formulate Eq. (\ref{eq:dev6}) as a function: $Z^x= f(\eta^x)-\xi$, $x$ is the iterator index. $\tau$ is the tolerance. $\tau$ is set as 0.001. }
		\STATE {\textbf{Input:} the start point $\eta^0$}
		\STATE {\textbf{Output:} final solution $\eta^*$ }
		\STATE {\textbf{repeat:}  }
		\STATE \hspace{0.2cm} {Calculate the gradient: $v^x =\bigtriangledown f(\eta^x) $; }
		\STATE \hspace{0.2cm} {Update the next step point: $\eta^{x+1} =\eta^{x}-f(\eta^x)/v^x $; }
		\STATE \hspace{0.2cm} {Calculate the function: $Z^{x+1}$; }
		\STATE {\textbf{until:} $|Z^x-Z^{x+1}|<\tau$ or $|Z^{x+1}|<\tau$} \\
	\end{algorithmic}
	\label{alg1}
\end{algorithm}

\subsection{The Overall Algorithm}
To refine the bit rate allocation, several issues need to be addressed. Firstly, as the parameters $k_{i,j}$ and $c_{i,j}$ can not be obtained before coding  current CTU, we utilize the parameters of  co-located CTU to estimate the bit rate of  current CTU. These parameters $c_{i,j}$ and $k_{i,j}$ are updated as:
\begin{equation}
\label{eq:kcupdated}
\begin{aligned}
c_{i,j}=\frac{r_{i,j}^{\rm actual} \lambda_{i,j}^{\rm actual}}{ d_{i,j}^{\rm actual}}, k_{i,j}=\frac{d_{i,j}^{\rm  actual} }{ \left ( r_{i,j}^{\rm actual} \right )^{c_{i,j}}  }, 	
\end{aligned}
\end{equation}
where $r_{i,j}^{\rm actual}$, $\lambda_{i,j}^{\rm actual} $, and $d_{i,j}^{\rm actual}$ are the actual coding results. 
$R_{i}$ is also updated by the actual texture bit rate $R_{i,j}^{\rm actual}$  of the $j$-th CTU  as:
\begin{equation}
\label{eq:Rupdated}
\begin{aligned}
R_{i}=R_{i}-R_{i,j}^{\rm actual}
\end{aligned}.
\end{equation}

\begin{figure}[t]
	\centering
	\includegraphics [width=0.5\textwidth] {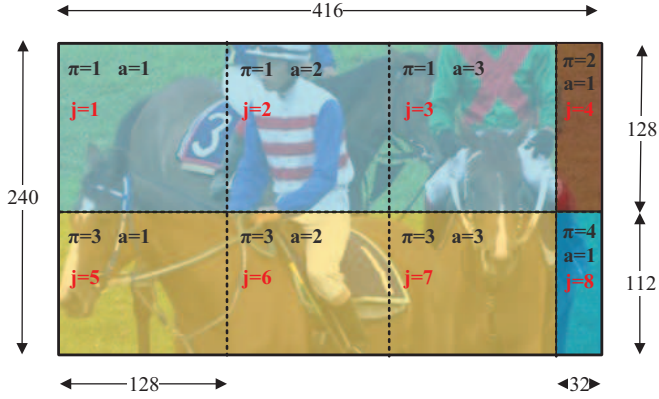}	
	\caption{The typical  effective CTU  sizes in the frame.}
	\label{partition}
\end{figure}

Secondly,  due to different video resolutions,  the Largest CU (LCU) size (128$\times$128) may lead to four different effective CTU  sizes (\textit{e.g.}, 128$\times$128, 32$\times$128, 128$\times$112, and 32$\times$112, as illustrated in Fig. \ref{partition}) of  CTUs in the frame. In general, larger effective CTU sizes need more bit rates to guarantee quality smoothness between different types of CTUs. Therefore, the effective CTU  sizes are also taken into consideration.   In the following paper, the $j$-th CTU is also denoted by $pi$ and $a$: it is also numbered as the $a$-th among all $pi$-th type of CTUs, as shown in Fig. \ref{partition}. 

To refine the bit rate allocation for different types of CTUs, the target bit rate of the $\pi$-th type  CTUs. $R_{i}^\pi$ in the $i$-th frame  is calculated by:
\begin{equation}
\label{eq:Rt}
\begin{aligned}
R_{i}^\pi=\frac{R_{i} \cdot M_{i}^\pi}{M_{i}^l},
\end{aligned}
\end{equation}
where $M_{i}^\pi$ and $M_{i}^l$ are the  pixel number of the $\pi$-th type of CTUs and  the number of  remaining pixels in  the frame, respectively. 

As analyzed above, the minimal utility  $u_{i,j}^0$ of the $j$-th CTU is also represented as $u_{i,j(\pi,a)}^0$. $u_{i,j}^0$ has been mentioned in Eq. (\ref{eq:aim}), which  regulates the minimum coding  quality for each CTU.  Therefore, the sum of bits corresponding to the uncoded $\pi$ type of CTUs   should  satisfy  $ {\textstyle \sum_{b = a}^{A_\pi}}  {  M_{i,j(\pi,b)}  r_{i,j(\pi,b)}^0}  \le R_{i}^\pi$. $A_\pi$ is the number of the uncoded $\pi$-th type  of CTUs   in the frame. $b$ is the index number. In this paper, we set a scale factor $S_{i,j}$, which is also represented as $S_{i,j(\pi,a)}$, to limit the minimum coding quality of the  $j$-th uncoded CTU \cite{GTAhmad2006,GTwang2013}.  $S_{i,j(\pi,a)}$ is computed as following:
\begin{equation}
\label{eq:S}
S_{i,j}= S_{i,j(\pi,a)}=   \frac{\varsigma R_{i}^\pi}{\sum_{b=a}^{A_\pi}\left (  M_{i,j(\pi,b)}\left (\frac{k_{i,j(\pi,b)}}{\tilde{d} _{i,j(\pi,b)} } \right)^{ \frac{1}{c_{i,j(\pi,b)}}} \right)},
\end{equation}
where $\varsigma$ is set as 0.7. $\tilde{d} _{i,j(\pi,b)} $ is the average  distortion of all co-located CTUs of previous frames at the same hierarchy of coding structure. In this paper, the adjusted minimal utility $u_{i,j(\pi,a)}^0$ is obtained by:
\begin{algorithm}[t]
	\caption{CTU Level Bit Rate Allocation Method}
	\begin{algorithmic} 
		\STATE {Initialize the parameters $k_{i,j}$ and $c_{i,j}$ }
		\STATE {\textbf{if}  the type of the current frame is intra \textbf{then}}
		\STATE \hspace{0.2cm} {Allocate the bit rate with the default VVC RC}
		\STATE {\textbf{else}  }
		\STATE \hspace{0.2cm}{\textbf{for}  all uncoded CTUs in a frame \textbf{do}}
		\STATE \hspace{0.75cm}{Calculate the minimal utility $u_{i,j}^0$ with Eqs.  (\ref{eq:S} and \ref{eq:u_ij0})}
		\STATE \hspace{0.5cm}{\textbf{if}  the distortion value of  correspongding CTU in the \\ \hspace{0.75cm} same frame level is equal to 0 \textbf{then}}		
		\STATE \hspace{0.75cm}{Obtain the estimated $\lambda_{i,j}$ with Eq. (\ref{eq:lambd})}
		\STATE \hspace{0.5cm}{\textbf{else}  }	
		\STATE \hspace{0.75cm}{Obtain the estimated $\lambda_{i,j}$ with Eq. (\ref{eq:lambda2})}	
		\STATE \hspace{0.5cm}{\textbf{end if}}				
		\STATE \hspace{0.5cm}{Adjust $\lambda_{i,j}$ with the estmated $\lambda$ value of the current frame  \\ \hspace{0.5cm} $\lambda^{\rm est}$ and the  $\lambda$ value of neighbor CTU $\lambda^{\rm nei}$ }
		\STATE \hspace{0.5cm}{\textbf{if} $\lambda^{\rm nei}>0$ \textbf{then}}
		\STATE \hspace{0.75cm}{  $\lambda_{i,j}=\max(\min(\lambda_{i,j},\lambda^{\rm nei} \cdot 2^{\frac{1}{3}}), \lambda^{\rm nei} \cdot 2^{\frac{-1}{3}} )$ }		
		\STATE \hspace{0.5cm}{\textbf{end if}}	
		\STATE \hspace{0.5cm}{\textbf{if} $\lambda^{\rm est}>0$ \textbf{then}}
		\STATE \hspace{0.75cm}{  $\lambda_{i,j}=\max(\min(\lambda_{i,j},\lambda^{\rm est} \cdot 2^{\frac{2}{3}}), \lambda^{\rm est} \cdot 2^{\frac{-2}{3}} )$ }	
		\STATE \hspace{0.5cm}{\textbf{else}  }		
		\STATE \hspace{0.75cm}{$\lambda_{i,j}=\max(\min(\lambda_{i,j},1000 \cdot 2^{4}), 10 \cdot 2^{4} )$ }		
		\STATE \hspace{0.5cm}{\textbf{end if}}		
		\STATE \hspace{0.5cm}{\textbf{if} $\lambda_{i,j}<0.1$ \textbf{then}}
		\STATE \hspace{0.75cm}{$\lambda_{i,j}=0.1$ }		
		\STATE \hspace{0.5cm}{\textbf{end if}}						
		\STATE \hspace{0.5cm}{Estimate the bpp  of  uncoded CTU $r_{i,j}$ by Eq. (\ref{eq:Restimated})}			
		\STATE \hspace{0.2cm}{\textbf{end for}}
		\STATE \hspace{0.2cm}{Calculate the refined estimated bit rate $R_{i,j}$ with Eq. (\ref{eq:Reij})}
		\STATE \hspace{0.2cm}{Encode the $j$-th CTU with the software}
		\STATE \hspace{0.2cm}{Obtain the actual bit rate $R_{i,j}^{\rm actual}$ and update the  remaining bit \\ \hspace{0.2cm} rates  $R_{i}$ with Eq. (\ref{eq:Rupdated})}
		\STATE \hspace{0.2cm}{Update the $k_{i,j}$ and $c_{i,j}$ parameters Eq. (\ref{eq:kcupdated})}

		\STATE { \textbf{end if}}		
	\end{algorithmic}
	\label{alg2}
\end{algorithm}

	\begin{table*}[]
		\caption{Comparison of RC performances in terms of the  Y-PSNR and RCError.}
		\label{LDB_YPSNR_RCError}
		\centering
		\scalebox{0.98}{
			\setlength{\tabcolsep}{1.25mm}{
				\begin{tabular}{|cc|cc|cc|cc|cc|cc|}
					\hline
					\multicolumn{1}{|c|}{\multirow{2}{*}{\textbf{Class}}} & \multirow{2}{*}{\textbf{Sequence}}    & \multicolumn{2}{c|}{\textbf{FixQP}}                                                                                                                  & \multicolumn{2}{c|}{\textbf{VTM13.0RC}}                                                                                                               & \multicolumn{2}{c|}{\textbf{SOSR'TBC21}}                                                                                                              & \multicolumn{2}{c|}{\textbf{MORC'TIP21}}                                                                                                              & \multicolumn{2}{c|}{\textbf{Proposed}}                                                                                                                \\ \cline{3-12} 
					\multicolumn{1}{|c|}{}                                &                                       & \multicolumn{1}{c|}{\textbf{\begin{tabular}[c]{@{}c@{}}Y-PSNR\\ (dB)\end{tabular}}} & \textbf{\begin{tabular}[c]{@{}c@{}}Rate\\ (kbps)\end{tabular}} & \multicolumn{1}{c|}{\textbf{\begin{tabular}[c]{@{}c@{}}Y-PSNR\\ (dB)\end{tabular}}} & \textbf{\begin{tabular}[c]{@{}c@{}}RCError\\ (\%)\end{tabular}} & \multicolumn{1}{c|}{\textbf{\begin{tabular}[c]{@{}c@{}}Y-PSNR\\ (dB)\end{tabular}}} & \textbf{\begin{tabular}[c]{@{}c@{}}RCError\\ (\%)\end{tabular}} & \multicolumn{1}{c|}{\textbf{\begin{tabular}[c]{@{}c@{}}Y-PSNR\\ (dB)\end{tabular}}} & \textbf{\begin{tabular}[c]{@{}c@{}}RCError\\ (\%)\end{tabular}} & \multicolumn{1}{c|}{\textbf{\begin{tabular}[c]{@{}c@{}}Y-PSNR\\ (dB)\end{tabular}}} & \textbf{\begin{tabular}[c]{@{}c@{}}RCError\\ (\%)\end{tabular}} \\ \hline
					\multicolumn{1}{|c|}{\multirow{5}{*}{\textbf{B}}}     & \textit{\textbf{MarketPlace}}         & \multicolumn{1}{c|}{37.2014}                                                        & 6216.883                                                       & \multicolumn{1}{c|}{37.5081}                                                        & 0.12                                                            & \multicolumn{1}{c|}{37.3057}                                                        & 0.12                                                            & \multicolumn{1}{c|}{37.4884}                                                        & 0.20                                                            & \multicolumn{1}{c|}{37.4935}                                                        & 0.12                                                            \\ 
					\multicolumn{1}{|c|}{}                                & \textit{\textbf{RitualDance}}         & \multicolumn{1}{c|}{39.4402}                                                        & 4652.504                                                       & \multicolumn{1}{c|}{39.4283}                                                        & 0.10                                                            & \multicolumn{1}{c|}{39.0506}                                                        & 0.09                                                            & \multicolumn{1}{c|}{39.4462}                                                        & 0.12                                                            & \multicolumn{1}{c|}{39.4089}                                                        & 0.10                                                            \\ 
					\multicolumn{1}{|c|}{}                                & \textit{\textbf{Cactus}}              & \multicolumn{1}{c|}{35.7330}                                                         & 5093.700                                                         & \multicolumn{1}{c|}{35.7530}                                                        & 0.10                                                            & \multicolumn{1}{c|}{35.5163}                                                        & 0.11                                                            & \multicolumn{1}{c|}{35.7329}                                                        & 0.26                                                            & \multicolumn{1}{c|}{35.7453}                                                        & 0.10                                                            \\  
					\multicolumn{1}{|c|}{}                                & \textit{\textbf{BasketballDrive}}     & \multicolumn{1}{c|}{36.7663}                                                        & 6136.736                                                       & \multicolumn{1}{c|}{36.6474}                                                        & 0.09                                                            & \multicolumn{1}{c|}{36.5220}                                                        & 0.09                                                            & \multicolumn{1}{c|}{36.6865}                                                        & 0.12                                                            & \multicolumn{1}{c|}{36.6792}                                                        & 0.09                                                            \\ 
					\multicolumn{1}{|c|}{}                                & \textit{\textbf{BQTerrace}}           & \multicolumn{1}{c|}{34.7598}                                                        & 13182.93                                                       & \multicolumn{1}{c|}{34.7343}                                                        & 0.18                                                            & \multicolumn{1}{c|}{34.3929}                                                        & 0.18                                                            & \multicolumn{1}{c|}{34.7524}                                                        & 0.19                                                            & \multicolumn{1}{c|}{34.7213}                                                        & 0.18                                                            \\ \hline
					\multicolumn{1}{|c|}{\multirow{4}{*}{\textbf{C}}}     & \textit{\textbf{BasketballDrill}}     & \multicolumn{1}{c|}{36.5083}                                                        & 1245.662                                                       & \multicolumn{1}{c|}{36.8879}                                                        & 0.29                                                            & \multicolumn{1}{c|}{36.8198}                                                        & 0.30                                                            & \multicolumn{1}{c|}{36.9374}                                                        & 0.32                                                            & \multicolumn{1}{c|}{36.8914}                                                        & 0.29                                                            \\  
					\multicolumn{1}{|c|}{}                                & \textit{\textbf{BQMall}}              & \multicolumn{1}{c|}{35.9642}                                                        & 1399.921                                                       & \multicolumn{1}{c|}{36.0326}                                                        & 0.31                                                            & \multicolumn{1}{c|}{35.8437}                                                        & 0.31                                                            & \multicolumn{1}{c|}{36.0213}                                                        & 0.34                                                            & \multicolumn{1}{c|}{36.0579}                                                        & 0.31                                                            \\ 
					\multicolumn{1}{|c|}{}                                & \textit{\textbf{PartyScene}}          & \multicolumn{1}{c|}{33.1892}                                                        & 2716.644                                                       & \multicolumn{1}{c|}{33.4161}                                                        & 0.17                                                            & \multicolumn{1}{c|}{33.3099}                                                        & 0.18                                                            & \multicolumn{1}{c|}{33.4067}                                                        & 0.33                                                            & \multicolumn{1}{c|}{33.3696}                                                        & 0.17                                                            \\  
					\multicolumn{1}{|c|}{}                                & \textit{\textbf{RaceHorses}}          & \multicolumn{1}{c|}{35.0175}                                                        & 1953.890                                                        & \multicolumn{1}{c|}{34.9717}                                                        & 0.14                                                            & \multicolumn{1}{c|}{34.7670}                                                        & 0.14                                                            & \multicolumn{1}{c|}{34.9307}                                                        & 0.16                                                            & \multicolumn{1}{c|}{34.9521}                                                        & 0.14                                                            \\ \hline
					\multicolumn{1}{|c|}{\multirow{4}{*}{\textbf{D}}}     & \textit{\textbf{BasketballPass}}      & \multicolumn{1}{c|}{35.7329}                                                        & 629.642                                                        & \multicolumn{1}{c|}{35.9812}                                                        & 0.53                                                            & \multicolumn{1}{c|}{35.8607}                                                        & 0.53                                                            & \multicolumn{1}{c|}{35.9573}                                                        & 0.56                                                            & \multicolumn{1}{c|}{35.9748}                                                        & 0.53                                                            \\  
					\multicolumn{1}{|c|}{}                                & \textit{\textbf{BQSquare}}            & \multicolumn{1}{c|}{33.2506}                                                        & 658.309                                                        & \multicolumn{1}{c|}{33.0898}                                                        & 0.97                                                            & \multicolumn{1}{c|}{32.9413}                                                        & 0.97                                                            & \multicolumn{1}{c|}{33.1068}                                                        & 1.04                                                            & \multicolumn{1}{c|}{33.1708}                                                        & 1.05                                                            \\
					\multicolumn{1}{|c|}{}                                & \textit{\textbf{BlowingBubbles}}      & \multicolumn{1}{c|}{32.9515}                                                        & 657.159                                                        & \multicolumn{1}{c|}{33.0947}                                                        & 0.64                                                            & \multicolumn{1}{c|}{33.0090}                                                        & 0.64                                                            & \multicolumn{1}{c|}{33.1026}                                                        & 0.75                                                            & \multicolumn{1}{c|}{33.1220}                                                        & 0.64                                                            \\ 
					\multicolumn{1}{|c|}{}                                & \textit{\textbf{RaceHorses}}          & \multicolumn{1}{c|}{34.1670}                                                         & 454.172                                                        & \multicolumn{1}{c|}{34.0853}                                                        & 0.48                                                            & \multicolumn{1}{c|}{33.9449}                                                        & 0.48                                                            & \multicolumn{1}{c|}{34.0435}                                                        & 0.49                                                            & \multicolumn{1}{c|}{34.0688}                                                        & 0.48                                                            \\ \hline
					\multicolumn{1}{|c|}{\multirow{3}{*}{\textbf{E}}}     & \textit{\textbf{FourPeople}}          & \multicolumn{1}{c|}{39.2264}                                                        & 690.026                                                        & \multicolumn{1}{c|}{39.1541}                                                        & 0.60                                                            & \multicolumn{1}{c|}{39.0818}                                                        & 0.61                                                            & \multicolumn{1}{c|}{39.1938}                                                        & 0.68                                                            & \multicolumn{1}{c|}{39.1793}                                                        & 0.59                                                            \\  
					\multicolumn{1}{|c|}{}                                & \textit{\textbf{Johnny}}              & \multicolumn{1}{c|}{40.4578}                                                        & 406.579                                                        & \multicolumn{1}{c|}{40.3865}                                                        & 1.29                                                            & \multicolumn{1}{c|}{40.2902}                                                        & 1.31                                                            & \multicolumn{1}{c|}{40.3978}                                                        & 1.53                                                            & \multicolumn{1}{c|}{40.3766}                                                        & 1.30                                                            \\  
					\multicolumn{1}{|c|}{}                                & \textit{\textbf{KristenAndSara}}      & \multicolumn{1}{c|}{40.3053}                                                        & 604.222                                                        & \multicolumn{1}{c|}{40.3089}                                                        & 0.77                                                            & \multicolumn{1}{c|}{40.1380}                                                        & 0.77                                                            & \multicolumn{1}{c|}{40.3171}                                                        & 0.82                                                            & \multicolumn{1}{c|}{40.3041}                                                        & 0.77                                                            \\ \hline
					\multicolumn{1}{|c|}{\multirow{4}{*}{\textbf{F}}}     & \textit{\textbf{BasketballDrillText}} & \multicolumn{1}{c|}{36.3085}                                                        & 1253.154                                                       & \multicolumn{1}{c|}{36.7391}                                                        & 0.29                                                            & \multicolumn{1}{c|}{36.6032}                                                        & 0.30                                                            & \multicolumn{1}{c|}{36.7617}                                                        & 0.32                                                            & \multicolumn{1}{c|}{36.7248}                                                        & 0.30                                                            \\  
					\multicolumn{1}{|c|}{}                                & \textit{\textbf{ArenaOfValor}}        & \multicolumn{1}{c|}{37.8757}                                                        & 5244.161                                                       & \multicolumn{1}{c|}{38.0913}                                                        & 0.10                                                            & \multicolumn{1}{c|}{38.1763}                                                        & 0.10                                                            & \multicolumn{1}{c|}{38.2164}                                                        & 0.11                                                            & \multicolumn{1}{c|}{38.1012}                                                        & 0.10                                                            \\ 
					\multicolumn{1}{|c|}{}                                & \textit{\textbf{SlideEditing}}        & \multicolumn{1}{c|}{41.6256}                                                        & 111.518                                                        & \multicolumn{1}{c|}{36.4628}                                                        & 5.69                                                            & \multicolumn{1}{c|}{37.9999}                                                        & 10.43                                                           & \multicolumn{1}{c|}{36.5385}                                                        & 7.85                                                            & \multicolumn{1}{c|}{36.6463}                                                        & 6.19                                                            \\  
					\multicolumn{1}{|c|}{}                                & \textit{\textbf{SlideShow}}           & \multicolumn{1}{c|}{43.9056}                                                        & 220.560                                                         & \multicolumn{1}{c|}{43.7232}                                                        & 14.30                                                           & \multicolumn{1}{c|}{43.8177}                                                        & 3.33                                                            & \multicolumn{1}{c|}{43.1223}                                                        & 8.41                                                            & \multicolumn{1}{c|}{44.1698}                                                        & 0.70                                                            \\ \hline
					\multicolumn{2}{|c|}{\textbf{Average}}                                                        & \multicolumn{1}{c|}{37.0193}                                                        & 2676.419                                                       & \multicolumn{1}{c|}{36.8248}                                                        & 1.36                                                            & \multicolumn{1}{c|}{36.7695}                                                        & 1.05                                                            & \multicolumn{1}{c|}{36.8080}                                                        & 1.23                                                            & \multicolumn{1}{c|}{\textbf{36.8579}}                                                        & \textbf{0.71}                                                            \\ \hline
				\end{tabular}
			}}
		\end{table*}
		
\begin{equation}
\label{eq:u_ij0}
\begin{aligned}
u_{i,j}^0=u_{i,j(\pi,a)}^0 = \frac{1}{\delta\tilde{d} _{i,j(\pi,a)} }  {\min\left(S_{i,j(\pi,a)},1\right)},	
\end{aligned}
\end{equation}
where the adjustment factor $\delta$ \cite{GTGAO2017} is defined as:
\begin{equation}
\label{eq:qpstep}
\begin{aligned}
\delta=\frac{\hat{Q}}{Q},	
\end{aligned}
\end{equation}
where $\hat{Q}$ is the estimated Quantization Step ($Q_{\rm  step}$) of the current frame. $Q$ is the $Q_{\rm step}$ of the previous frame in the same frame level.

Finally, by substituting Eq. (\ref{eq:lambda2}) into Eq. (\ref{eq:R}), we estimate the target bpp $r_{i,j}$ of the $j$-th CTU as:
\begin{equation}
\label{eq:Restimated}
r_{i,j}= \left ( \frac{k_{i,j} u_{i,j}^0\eta^*}{c_{i,j}+\eta^*} \right ) ^{\frac{1}{c_{i,j}}}.
\end{equation}
To optimize the overall coding quality in the frame, we estimate the bit rates for all uncoded CTUs and adopt  sliding windows \cite{slidewindowsLi2014} to refine the estimated bit rate $R_{i,j}$. $R_{i,j}$ is calculated by:
\begin{equation}
\label{eq:Reij}
\begin{aligned}
&R_{i,j} = \lfloor M_{i,j} r_{i,j}-\frac{1}{W} \left (\sum \limits_{j=1}^{N_i} M_{i,j} r_{i,j}   -R_{i}  \right )  +0.5 \rfloor,
\end{aligned}
\end{equation}
where $W$ is the size of sliding windows, which is set as the minimum value of 4 and $N_i$.

In summary, the detailed bit rate allocation method is shown in Algorithm \ref{alg2}.

\section{Experiment}
\label{experiment}

		\subsection{Simulation Setup}
		The proposed method is implemented on the VVC reference software VTM13.0 \cite{VVC}. All experimental results are regulated under the Common Test Condition (CTC) document JVET-T2010 \cite{CTC-F.Bossen20} to verify the effectiveness of the proposed method.   Four target bit rates are generated with four fixed Qps: 22, 27, 32, and 37, respectively, by VTM13.0 without RC (denoted as FixQP). For comparison, we  obtain the performance of the up-to-date CTU-level bit rate allocation methods \cite{TCSVT21-Liu,TBC21-Li} on the same software platform-VTM13.0. They are denoted as  SOSR'TBC21 \cite{TBC21-Li} and MORC'TIP21 \cite{TCSVT21-Liu}, respectively. Since all compared methods were implemented on Low-Delay B (LD) configuration, we also use this configuration for fair comparison. 
		
		\subsection{Comparison on Bit  Rate Allocation}
		The luminance component of PSNR (Y-PSNR) and RCError metrics are used to evaluate the method's performance.  Y-PSNR  is positively correlated with the visual quality of  compressed videos. RCError indicates the inaccuracy of bitrate allocation.  A higher RCError indicates that the method has a lower bit rate allocation accuracy, and vice versa. It is calculated by:
		\begin{equation}
		\begin{aligned}
		RCError=\frac{\left | R_{\rm target}-R_{\rm actual} \right | }{R_{\rm target} }\times 100\% ,	
		\end{aligned}
		\end{equation}
		where $R_{\rm target}$ is the target bit rate. $R_{\rm actual}$ is the actual coding bit rate of the proposed method.
		
In Table \ref{LDB_YPSNR_RCError}, the default RC scheme of  VTM13.0 (VTM13.0RC) achieves a relatively acceptable visual performance with an average Y-PSNR of 36.8248 dB and an average RCError of 1.36\%.   SOSR'TBC21 and MORC'TIP21  achieve  lower coding quality with  36.7695dB and 36.8080dB, respectively. Nevertheless, they perform better than  VTM13.0RC   in the RCError metric with  0.31\% and 0.13\%. As for  the proposed method, its average Y-PSNR  difference from FixQP is only 0.1614 dB. Comparing the performance of  VTM13.0RC, SOSR'TBC21, and MORC'TIP21  in Y-PSNR and RCError, we can find that the proposed method achieves an average Y-PSNR of 36.8579dB and an average RCError of 0.71\%. It outperforms  the performances of benchmarks. Especially for the RCError metric, the proposed method reduces the average bit rate accuracy of  VTM13.0RC, SOSR'TBC21, and MORC'TIP21  by about 47.8\%, 32.4\%, and 42.3\%, respectively.
	\begin{table*}[]
		\centering
		\caption{Comparison of RD performances in terms of BDPSNR and BDBR.}
		\label{tab:BDBRLDB}
		\scalebox{0.98}{
			\setlength{\tabcolsep}{0.8mm}{
				\begin{tabular}{|cc|cc|cc|cc|cc|}
					\hline
					\multicolumn{1}{|c|}{\multirow{3}{*}{\textbf{Class}}} & \multirow{3}{*}{\textbf{Sequence}}    & \multicolumn{2}{c|}{\textbf{VTM13.0RC}}                                                          & \multicolumn{2}{c|}{\textbf{SOSR'TBC21}}                                                         & \multicolumn{2}{c|}{\textbf{MORC'TIP21}}                                                         & \multicolumn{2}{c|}{\textbf{Proposed}}                                                           \\ \cline{3-10} 
					\multicolumn{1}{|c|}{}                                &                                       & \multicolumn{1}{c|}{\multirow{2}{*}{\textbf{BDPSNR (dB)}}} & \multirow{2}{*}{\textbf{BDBR (\%)}} & \multicolumn{1}{c|}{\multirow{2}{*}{\textbf{BDPSNR (dB)}}} & \multirow{2}{*}{\textbf{BDBR (\%)}} & \multicolumn{1}{c|}{\multirow{2}{*}{\textbf{BDPSNR (dB)}}} & \multirow{2}{*}{\textbf{BDBR (\%)}} & \multicolumn{1}{c|}{\multirow{2}{*}{\textbf{BDPSNR (dB)}}} & \multirow{2}{*}{\textbf{BDBR (\%)}} \\
					\multicolumn{1}{|c|}{}                                &                                       & \multicolumn{1}{c|}{}                                      &                                     & \multicolumn{1}{c|}{}                                      &                                     & \multicolumn{1}{c|}{}                                      &                                     & \multicolumn{1}{c|}{}                                      &                                     \\ \hline
					\multicolumn{1}{|c|}{\multirow{5}{*}{\textbf{B}}}     & \textit{\textbf{MarketPlace}}         & \multicolumn{1}{c|}{0.2913}                                & -11.32                              & \multicolumn{1}{c|}{0.0870}                                & -4.32                               & \multicolumn{1}{c|}{0.2702}                                & -10.64                              & \multicolumn{1}{c|}{0.2769}                                & -10.92                              \\  
					\multicolumn{1}{|c|}{}                                & \textit{\textbf{RitualDance}}         & \multicolumn{1}{c|}{-0.0180}                               & 0.38                                & \multicolumn{1}{c|}{-0.4231}                               & 9.45                                & \multicolumn{1}{c|}{0.0029}                                & -0.06                               & \multicolumn{1}{c|}{-0.0437}                               & 0.92                                \\ 
					\multicolumn{1}{|c|}{}                                & \textit{\textbf{Cactus}}              & \multicolumn{1}{c|}{0.0085}                                & 0.23                                & \multicolumn{1}{c|}{-0.2095}                               & 10.75                               & \multicolumn{1}{c|}{-0.0147}                               & 1.25                                & \multicolumn{1}{c|}{-0.0177}                               & 1.21                                \\ 
					\multicolumn{1}{|c|}{}                                & \textit{\textbf{BasketballDrive}}     & \multicolumn{1}{c|}{-0.1148}                               & 5.31                                & \multicolumn{1}{c|}{-0.2405}                               & 11.19                               & \multicolumn{1}{c|}{-0.0707}                               & 3.34                                & \multicolumn{1}{c|}{-0.0797}                               & 3.74                                \\ 
					\multicolumn{1}{|c|}{}                                & \textit{\textbf{BQTerrace}}           & \multicolumn{1}{c|}{-0.0504}                               & 3.66                                & \multicolumn{1}{c|}{-0.3892}                               & 30.96                               & \multicolumn{1}{c|}{-0.0447}                               & 3.24                                & \multicolumn{1}{c|}{-0.0853}                               & 5.74                                \\ \hline
					\multicolumn{1}{|c|}{\multirow{4}{*}{\textbf{C}}}     & \textit{\textbf{BasketballDrill}}     & \multicolumn{1}{c|}{0.3269}                                & -7.88                               & \multicolumn{1}{c|}{0.2688}                                & -6.51                               & \multicolumn{1}{c|}{0.3819}                                & -9.16                               & \multicolumn{1}{c|}{0.3360}                                & -8.13                               \\
					\multicolumn{1}{|c|}{}                                & \textit{\textbf{BQMall}}              & \multicolumn{1}{c|}{0.0554}                                & -1.46                               & \multicolumn{1}{c|}{-0.1329}                               & 3.56                                & \multicolumn{1}{c|}{0.0442}                                & -1.16                               & \multicolumn{1}{c|}{0.0818}                                & -2.15                               \\  
					\multicolumn{1}{|c|}{}                                & \textit{\textbf{PartyScene}}          & \multicolumn{1}{c|}{0.1798}                                & -4.43                               & \multicolumn{1}{c|}{0.0684}                                & -1.69                               & \multicolumn{1}{c|}{0.1592}                                & -3.95                               & \multicolumn{1}{c|}{0.1339}                                & -3.32                               \\  
					\multicolumn{1}{|c|}{}                                & \textit{\textbf{RaceHorses}}          & \multicolumn{1}{c|}{-0.0201}                               & 0.54                                & \multicolumn{1}{c|}{-0.2236}                               & 6.22                                & \multicolumn{1}{c|}{-0.0647}                               & 1.76                                & \multicolumn{1}{c|}{-0.0404}                               & 1.09                                \\ \hline
					\multicolumn{1}{|c|}{\multirow{4}{*}{\textbf{D}}}     & \textit{\textbf{BasketballPass}}      & \multicolumn{1}{c|}{0.2098}                                & -4.16                               & \multicolumn{1}{c|}{0.0922}                                & -1.86                               & \multicolumn{1}{c|}{0.1834}                                & -3.65                               & \multicolumn{1}{c|}{0.2031}                                & -4.06                               \\ 
					\multicolumn{1}{|c|}{}                                & \textit{\textbf{BQSquare}}            & \multicolumn{1}{c|}{-0.1394}                               & 3.73                                & \multicolumn{1}{c|}{-0.3658}                               & 11.91                               & \multicolumn{1}{c|}{-0.1868}                               & 5.85                                & \multicolumn{1}{c|}{-0.1194}                               & 3.75                                \\  
					\multicolumn{1}{|c|}{}                                & \textit{\textbf{BlowingBubbles}}      & \multicolumn{1}{c|}{0.1200}                                & -3.07                               & \multicolumn{1}{c|}{0.0325}                                & -0.80                               & \multicolumn{1}{c|}{0.1239}                                & -3.15                               & \multicolumn{1}{c|}{0.1377}                                & -3.50                               \\ 
					\multicolumn{1}{|c|}{}                                & \textit{\textbf{RaceHorses}}          & \multicolumn{1}{c|}{-0.0932}                               & 2.13                                & \multicolumn{1}{c|}{-0.2215}                               & 4.99                                & \multicolumn{1}{c|}{-0.1378}                               & 3.10                                & \multicolumn{1}{c|}{-0.1101}                               & 2.51                                \\ \hline
					\multicolumn{1}{|c|}{\multirow{3}{*}{\textbf{E}}}     & \textit{\textbf{FourPeople}}          & \multicolumn{1}{c|}{-0.1276}                               & 4.16                                & \multicolumn{1}{c|}{-0.1720}                               & 5.66                                & \multicolumn{1}{c|}{-0.0575}                               & 2.24                                & \multicolumn{1}{c|}{-0.0869}                               & 3.10                                \\
					\multicolumn{1}{|c|}{}                                & \textit{\textbf{Johnny}}              & \multicolumn{1}{c|}{-0.0799}                               & 4.88                                & \multicolumn{1}{c|}{-0.2168}                               & 10.68                               & \multicolumn{1}{c|}{-0.0936}                               & 4.65                                & \multicolumn{1}{c|}{-0.0927}                               & 5.05                                \\  
					\multicolumn{1}{|c|}{}                                & \textit{\textbf{KristenAndSara}}      & \multicolumn{1}{c|}{-0.0326}                               & 1.60                                & \multicolumn{1}{c|}{-0.1817}                               & 7.83                                & \multicolumn{1}{c|}{-0.0051}                               & 0.75                                & \multicolumn{1}{c|}{-0.0334}                               & 1.65                                \\ \hline
					\multicolumn{1}{|c|}{\multirow{4}{*}{\textbf{F}}}     & \textit{\textbf{BasketballDrillText}} & \multicolumn{1}{c|}{0.3749}                                & -8.60                               & \multicolumn{1}{c|}{0.2553}                                & -5.86                               & \multicolumn{1}{c|}{0.4041}                                & -9.20                               & \multicolumn{1}{c|}{0.3696}                                & -8.47                               \\  
					\multicolumn{1}{|c|}{}                                & \textit{\textbf{ArenaOfValor}}        & \multicolumn{1}{c|}{0.1669}                                & -4.09                               & \multicolumn{1}{c|}{0.2524}                                & -6.22                               & \multicolumn{1}{c|}{0.2986}                                & -7.41                               & \multicolumn{1}{c|}{0.1761}                                & -4.35                               \\ 
					\multicolumn{1}{|c|}{}                                & \textit{\textbf{SlideEditing}}        & \multicolumn{1}{c|}{-6.0720}                               & 50.90                               & \multicolumn{1}{c|}{-5.4625}                               & 46.26                               & \multicolumn{1}{c|}{-6.3164}                               & 53.22                               & \multicolumn{1}{c|}{-6.1526}                               & 49.96                               \\  
					\multicolumn{1}{|c|}{}                                & \textit{\textbf{SlideShow}}           & \multicolumn{1}{c|}{-0.1326}                               & 5.93                                & \multicolumn{1}{c|}{-0.0566}                               & 0.86                                & \multicolumn{1}{c|}{-0.4627}                               & 10.29                               & \multicolumn{1}{c|}{0.3659}                                & -4.65                               \\ \hline
					\multicolumn{2}{|c|}{\textbf{Average}}                                                        & \multicolumn{1}{c|}{-0.2574}                               & 1.92                                & \multicolumn{1}{c|}{-0.3620}                               & 6.65                                & \multicolumn{1}{c|}{-0.2793}                               & 2.07                                & \multicolumn{1}{c|}{\textbf{-0.2391}}                               & \textbf{1.46}                                \\ \hline
				\end{tabular}}}
			\end{table*}	
 		\subsection{Comparison on RD Performances }
 		The RD performance of the proposed method is evaluated with Bjontegaard Average Peak-Signal-to-Noise-Ratio (BDPSNR)    and Bjontegaard Average Bit Rate (BDBR) \cite{BDBR-G.Bjontegaard01}. These metrics are calculated by Y-PSNR and bit rate.  BDPSNR  implies the average coding quality in dB with the same bit rate.  
 		The other metric, BDBR , implies the average increment of bit rate in percentage with the same visual quality. In other words, the RD performance of a method shows a positive correlation with  BDPSNR  and a negative correlation with  BDBR.
 		
 		This paper evaluates  BDPSNR  and BDBR  by comparing the methods with  FixQP.  As shown in Table \ref{tab:BDBRLDB},  SOSR'TBC21   achieves an average BDPSNR  of -0.3620dB  and  an average BDBR of 6.65\%. As the method is designed to enhance the visual quality  in HEVC and  the bit allocation strategy for all uncoded CTUs is not designed, the performance of SOSR'TBC21 does not achieve the best performance. MORC'TIP21 achieves a competive performance with an average BDPSNR  of -0.2793dB and   an average BDBR of 2.07\%. Although  MORC'TIP21  is a CTU-level bit rate allocation scheme, the allocation scheme allocates  bit rates  once for all.  In contrast, our bit rate allocation scheme of the proposed method  is dynamic and adaptive to complex video contents.  The proposed method achieves   the highest average BDPSNR of -0.2357dB and the lowest average BDBR  of 1.44\%. In Table \ref{tab:BDBRLDB}, the proposed method is  superior to  VTM13.0RC, SOSR'TBC21, and MORC'TIP21  in the average BDPSNR. In addition,  compared with  VTM13.0RC, SOSR'TBC21, and MORC'TIP21, the proposed method enhances the average BDBR by  0.48\%, 5.21\%, and 0.63\%, respectively.  
 
 In Table \ref{tab:BDBRLDB}, we can also find that most sequences (15/20) achieve optimal or sub-optimal BDPSNR and  BDBR performances. The performances  derive from the fact that the estimated target bit rate of  current CTU is obtained with the average previous CTU distortion. It also verifies that game theory can well balance  bit rates for uncoded CTUs, which can avoid the large-scale quality fluctuations.  In addition, the performance of  \textit{Slideshow} sequence indicates that the proposed method has significant advantages for the scene where the space complexity of the screen content fluctuates greatly. This benefits from the advantages of the game theory-based model. The greater  difference between the utility functions of the players indicates the better  global performance of the uncertain information game.

		\begin{table}[]
			
			\caption{Video quality and bit rate fluctuation}
			\centering
			\label{tab:vqrf}
			\scalebox{0.95}{
				\setlength{\tabcolsep}{0.9mm}{
					\begin{tabular}{|c|cccc|}
						\hline
						
						\textbf{Method}
						& \multicolumn{1}{c|}{\textbf{VTM13.0RC}} & \multicolumn{1}{c|}{\textbf{SOSR'TBC21}}  & \multicolumn{1}{c|}{\textbf{MORC'TIP21}}  & \textbf{Proposed} 
						
						\\ \hline
						\textbf{Average}	& \multicolumn{4}{c|}{\textbf{Rate}}                                                                   \\ \hline
						\textbf{Var}                & \multicolumn{1}{c|}{1.29e+10}  & \multicolumn{1}{c|}{1.31e+10} & \multicolumn{1}{c|}{1.66e+10} & \textbf{1.28e+10} \\ \hline
						\textbf{Mean}               & \multicolumn{1}{c|}{50229}     & \multicolumn{1}{c|}{50095}    & \multicolumn{1}{c|}{50191}    & \textbf{50080}    \\ \hline				
						\textbf{Average}                  & \multicolumn{4}{c|}{\textbf{Y-PSNR}}    
						\\ \hline
						\textbf{Var}                & \multicolumn{1}{c|}{7.7934}    & \multicolumn{1}{c|}{7.36424}  & \multicolumn{1}{c|}{\textbf{6.5994}}   & 7.1703   \\ \hline
						
						\textbf{Mean}               & \multicolumn{1}{c|}{36.8248}   & \multicolumn{1}{c|}{36.7695}  & \multicolumn{1}{c|}{36.8080}  & \textbf{36.8579}  \\ \hline                                                                    
					\end{tabular}}}
				\end{table}

	\begin{figure*}[t]
		\centering

		\begin {minipage} [htbp] {0.495\textwidth}
		\centering
		\includegraphics [width=1.0\textwidth] {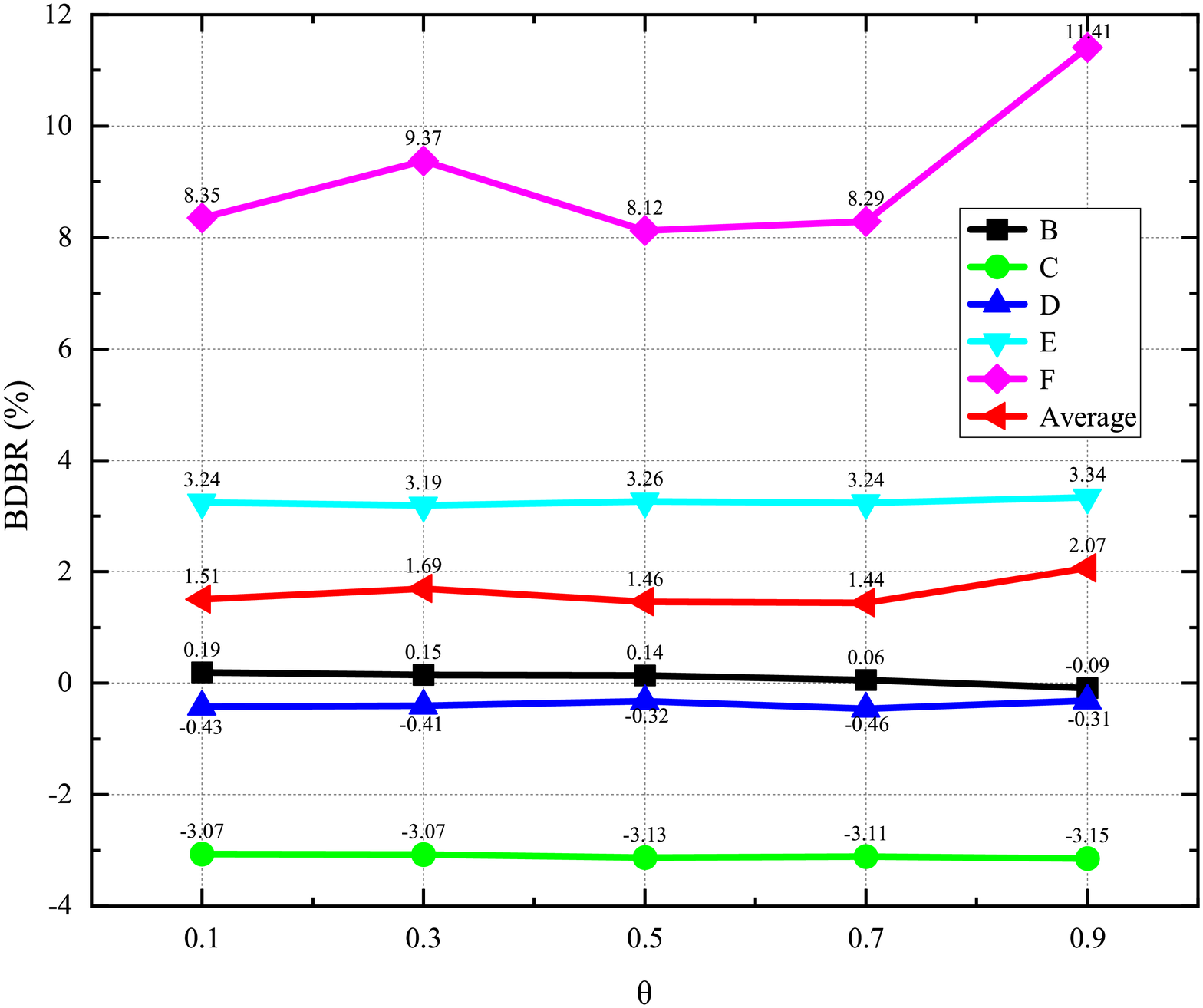}
		\centerline{(a) BDBR-$\varsigma$}
		\end {minipage}
		\begin {minipage} [htbp] {0.495\textwidth}
		\centering
		\includegraphics [width=1.0\textwidth] {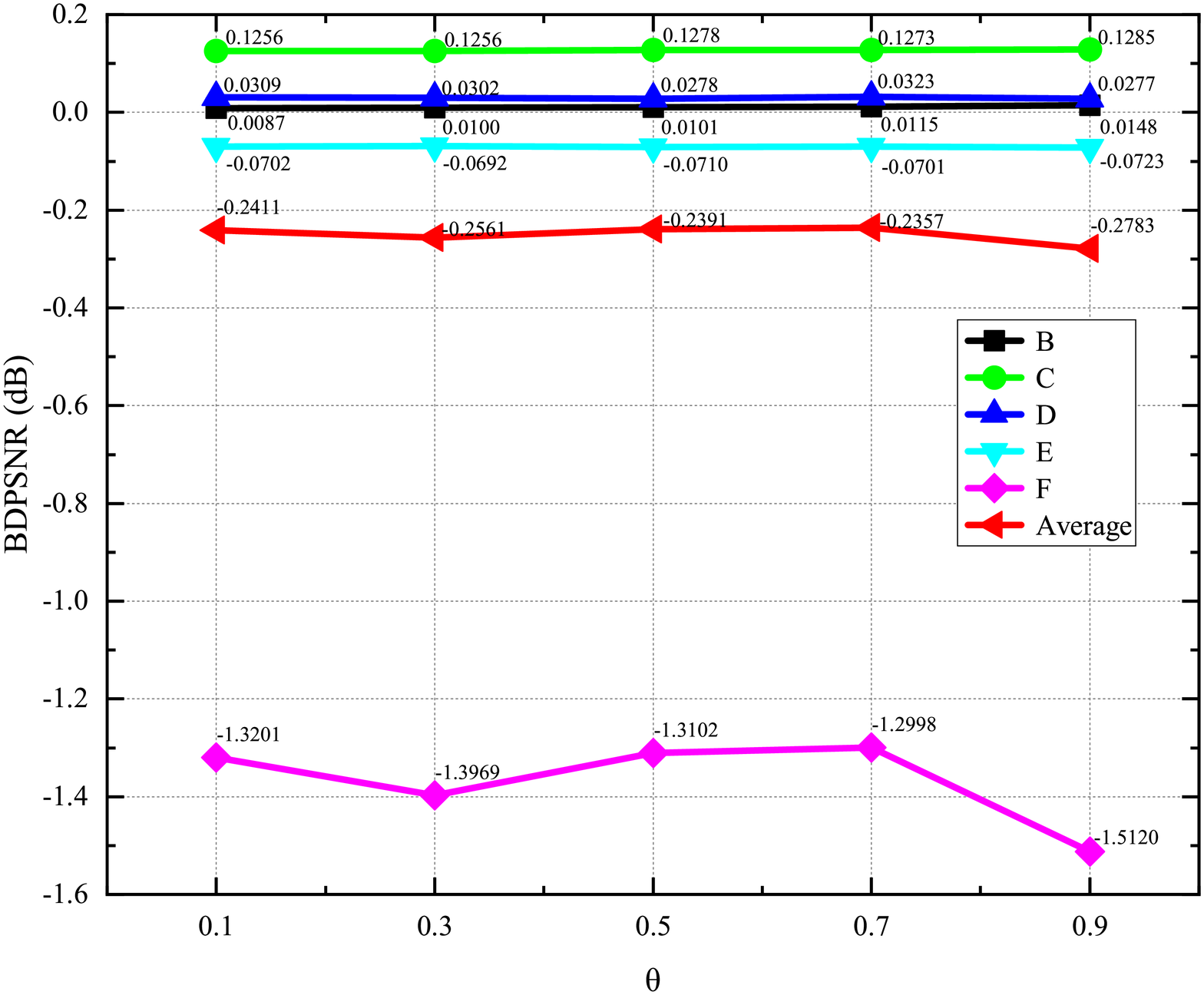}
		\centerline{(b) BDPSNR-$\varsigma$}
		\end {minipage}			
		\caption{ Performance of the proposed method with different $\varsigma$ values.}
				\label{tiaocan}
	\end{figure*}

\subsection{Video Quality and Bit Rate Fluctuation}
Variance and mean values are also used to evaluate the video quality and bit rate fluctuation of the sequence. A lower variance value indicates a lower fluctuation, and vice versa. A higher mean value implies a better performance of algorithm.   The average variance and mean values of all mandatory sequences  in Table \ref{tab:vqrf} indicate the proposed method achieves the lowest bit rate fluctuation. This achievement may be attributed to the adaptive CTU-level bit allocation in our method. It also implies that the proposed method is more suitable for complex video transmission environments. For the average mean of Y-PSNR, our proposed method achieves the optimal performance. Although the proposed method is not designed to control the video quality fluctuation, it also outperforms  VTM13.ORC and SOSR'TBC21. As MORC'TIP21  is proposed to control the video quality fluctuation, it achieves the best average variance of Y-PSNR. However, it also leads to the highest average variance of bit rate. Compared with  MORC'TIP21, the proposed method can better balance the fluctuation between  Y-PSNR and bit rate.

\begin{table}[]
	\centering
	\caption{Computational overhead.}
	\label{overhead}
	\scalebox{0.8}{
		\setlength{\tabcolsep}{1mm}{
\begin{tabular}{|c|cc|ccc|}
	\hline
	\multirow{2}{*}{-}              & \multicolumn{1}{c|}{\multirow{2}{*}{\textbf{Qp}}} & \textbf{Time cost (s)} & \multicolumn{3}{c|}{\textbf{Overhead(\%)}}                                                        \\ \cline{3-6} 
	& \multicolumn{1}{c|}{}                             & \textbf{VTM13.0RC}     & \multicolumn{1}{c|}{\textbf{SOSR'TBC21}} & \multicolumn{1}{c|}{\textbf{MORC'TIP21}} & \textbf{Proposed} \\ \hline
	\multirow{4}{*}{\textbf{\textit{\begin{tabular}[c]{@{}c@{}}Cactus \\ (Class B)\end
					{tabular}}}}       & \multicolumn{1}{c|}{22}                           & 187659                 & \multicolumn{1}{c|}{0.35}             & \multicolumn{1}{c|}{4.58}             & 0.02              \\
	 
	& \multicolumn{1}{c|}{27}                           & 109539                 & \multicolumn{1}{c|}{2.04}             & \multicolumn{1}{c|}{4.14}             & 0.68              \\  
	& \multicolumn{1}{c|}{32}                           & 70273                  & \multicolumn{1}{c|}{0.34}             & \multicolumn{1}{c|}{3.78}             & 1.93              \\ 
	& \multicolumn{1}{c|}{37}                           & 44224                  & \multicolumn{1}{c|}{0.60}             & \multicolumn{1}{c|}{3.73}             & 0.40              \\ \hline
	\multirow{4}{*}{\textbf{\textit{\begin{tabular}[c]{@{}c@{}} RaceHorses\\ (Class C)\end
					{tabular}}}}     & \multicolumn{1}{c|}{22}                           & 35785                  & \multicolumn{1}{c|}{-0.25}            & \multicolumn{1}{c|}{0.83}             & -0.08             \\ 
	& \multicolumn{1}{c|}{27}                           & 25723                  & \multicolumn{1}{c|}{0.72}             & \multicolumn{1}{c|}{0.24}             & 0.07              \\ 
	& \multicolumn{1}{c|}{32}                           & 17559                  & \multicolumn{1}{c|}{-0.72}            & \multicolumn{1}{c|}{-0.28}            & -0.70             \\  
	& \multicolumn{1}{c|}{37}                           & 10932                  & \multicolumn{1}{c|}{-2.13}            & \multicolumn{1}{c|}{-0.41}            & 0.41              \\ \hline
	\multirow{4}{*}{\textbf{\textit{\begin{tabular}[c]{@{}c@{}} BlowingBubbles\\ \centering (Class D)\end
					{tabular}}}} & \multicolumn{1}{c|}{22}                           & 11793                  & \multicolumn{1}{c|}{-3.18}            & \multicolumn{1}{c|}{3.01}             & -1.59             \\  
	& \multicolumn{1}{c|}{27}                           & 9098                   & \multicolumn{1}{c|}{-3.90}            & \multicolumn{1}{c|}{-4.24}            & -3.17             \\  
	& \multicolumn{1}{c|}{32}                           & 5506                   & \multicolumn{1}{c|}{-5.61}            & \multicolumn{1}{c|}{-1.19}            & -1.63             \\ 
	& \multicolumn{1}{c|}{37}                           & 3410                   & \multicolumn{1}{c|}{-4.49}            & \multicolumn{1}{c|}{-3.17}            & -2.17             \\ \hline
	\multirow{4}{*}{\textbf{\textit{\begin{tabular}[c]{@{}c@{}} Johnny \\(Class E)\end
					{tabular}}}}        & \multicolumn{1}{c|}{22}                           & 31898                  & \multicolumn{1}{c|}{4.98}             & \multicolumn{1}{c|}{2.42}             & 1.23              \\ 
	& \multicolumn{1}{c|}{27}                           & 17577                  & \multicolumn{1}{c|}{1.43}             & \multicolumn{1}{c|}{1.44}             & 2.08              \\ 
	& \multicolumn{1}{c|}{32}                           & 11208                  & \multicolumn{1}{c|}{5.86}             & \multicolumn{1}{c|}{-1.15}            & 1.43              \\  
	& \multicolumn{1}{c|}{37}                           & 7323                   & \multicolumn{1}{c|}{9.72}             & \multicolumn{1}{c|}{-0.77}            & 2.21              \\ \hline
	Average                         & \multicolumn{2}{c|}{-}                                                     & \multicolumn{1}{c|}{0.36}             & \multicolumn{1}{c|}{0.81}             & \textbf{0.07}              \\ \hline
\end{tabular}}}
		\end{table}

\subsection{Parameter Sensitivity Analysis}
Parameter sensitivity analysis is conducted for different $\varsigma$ values of Eq. (\ref{eq:S}) to investigate the performance variations of the proposed method. The parameter $\varsigma$ is the ratio between the guaranteed minimum utility bit rate and the  target bit rate. Here, we examine this parameter from  0.1 to 0.9 with a step size of 0.2. This experiment is conducted on the mandatory test sequences of the CTC. The  BDBR and BDPSNR performance of these parameters are shown in Fig. \ref{tiaocan}, where the BDBR and BDPSNR values corresponding to Class "B", "C", "D", "E", and "F" represent the average values of all sequences of the corresponding category in the CTC document. "Average" represents the average value of all test sequences.  From the figure, the proposed method achieves the optimal performance with an average BDBR of 1.44\%  and an average BDPSNR of -0.2357dB  when the value of $\varsigma$ parameter is 0.7. 

\subsection{Computational Overhead}	
We select four typical sequences (\textit{Cactus, RaceHorses, BlowingBubbles, and Johnny}) to evaluate the overheads of the methods. The experiments are carried out on the computer with @ 3.80GHz, 3.79GHz processor, and 16GB of RAM, running Windows 10. We compare the overhead of SOSR'TBC21, MORC'TIP21, and the proposed method with VTM13.0RC. The overhead $\Delta T$ is calculated by: 
		\begin{equation}
		\Delta T=\frac{  T_{\rm{method}}-T_{\rm{org}}   }{T_{\rm{org}} }\times 100\% ,\\	
		\end{equation}
where $T_{\rm method}$ represents the time cost of  a compared method. $T_{\rm org}$ is the time cost of  VTM13.0RC, which is set as the baseline. From Table \ref{overhead}, the overheads of  SOSR'TBC21, MORC'TIP21, and the proposed method are 0.36\% , 0.81\%, and 0.07\%, respectively. It shows that the overhead of the proposed method is lower than the overheads of  SOSR'TBC21 and MORC'TIP21.  Therefore, our method achieves a high RC performance with a negligible computational overhead.

	\section{Conclusion}
	\label{Conclusion}
This paper proposes a $\lambda$-domain RC based on game theory to enhance the RD performance of video coding. We first formulate  bit rate allocation  with  $\lambda$-domain RD model and game theory. Then, we obtain the optimal $\lambda$ value of uncoded CTU  and use it to estimate the bit rates of the next uncoded CTU. In addition, the overall RC  framework  is also designed. Experimental results show that the proposed method enhances a significant bit rate accuracy and obtains superior BDBR and BDPSNR performances, whilst maintaining a  negligible computational overhead.
\bibliographystyle{ACM-Reference-Format}
\bibliography{GT2022}


\begin{thebibliography}{30}


\ifx \showCODEN    \undefined \def \showCODEN     #1{\unskip}     \fi
\ifx \showDOI      \undefined \def \showDOI       #1{#1}\fi
\ifx \showISBNx    \undefined \def \showISBNx     #1{\unskip}     \fi
\ifx \showISBNxiii \undefined \def \showISBNxiii  #1{\unskip}     \fi
\ifx \showISSN     \undefined \def \showISSN      #1{\unskip}     \fi
\ifx \showLCCN     \undefined \def \showLCCN      #1{\unskip}     \fi
\ifx \shownote     \undefined \def \shownote      #1{#1}          \fi
\ifx \showarticletitle \undefined \def \showarticletitle #1{#1}   \fi
\ifx \showURL      \undefined \def \showURL       {\relax}        \fi
\providecommand\bibfield[2]{#2}
\providecommand\bibinfo[2]{#2}
\providecommand\natexlab[1]{#1}
\providecommand\showeprint[2][]{arXiv:#2}

\bibitem[\protect\citeauthoryear{Ahmad and Luo}{Ahmad and Luo}{2006}]%
        {GTAhmad2006}
\bibfield{author}{\bibinfo{person}{I. Ahmad} {and} \bibinfo{person}{Jiancong
  Luo}.} \bibinfo{year}{2006}\natexlab{}.
\newblock \showarticletitle{On using game theory to optimize the rate control
  in video coding}.
\newblock \bibinfo{journal}{\emph{IEEE Transactions on Circuits and Systems for
  Video Technology}} \bibinfo{volume}{16}, \bibinfo{number}{2}
  (\bibinfo{year}{2006}), \bibinfo{pages}{209--219}.
\newblock
\urldef\tempurl%
\url{https://doi.org/10.1109/TCSVT.2005.856899}
\showDOI{\tempurl}


\bibitem[\protect\citeauthoryear{Bjontegaard}{Bjontegaard}{2001}]%
        {BDBR-G.Bjontegaard01}
\bibfield{author}{\bibinfo{person}{Gisle Bjontegaard}.}
  \bibinfo{year}{2001}\natexlab{}.
\newblock \showarticletitle{Calculation of average PSNR differences between
  RD-curves}.
\newblock \bibinfo{journal}{\emph{VCEG-M33}} (\bibinfo{year}{2001}).
\newblock


\bibitem[\protect\citeauthoryear{{Bossen}, {Boyce}, {Suehring}, {Li}, and
  {Seregin}}{{Bossen} et~al\mbox{.}}{2020}]%
        {CTC-F.Bossen20}
\bibfield{author}{\bibinfo{person}{Frank {Bossen}}, \bibinfo{person}{Jill
  {Boyce}}, \bibinfo{person}{Karsten {Suehring}}, \bibinfo{person}{Xiang {Li}},
  {and} \bibinfo{person}{Vadim {Seregin}}.} \bibinfo{year}{2020}\natexlab{}.
\newblock \showarticletitle{VTM common test conditions and software reference
  configurations for SDR video}.
\newblock \bibinfo{journal}{\emph{JVET document, JVET-T2010}}
  (\bibinfo{year}{2020}).
\newblock


\bibitem[\protect\citeauthoryear{{Bossen}, {Li}, {Suehring}, {Seregin}, and
  {Tourapis}}{{Bossen} et~al\mbox{.}}{2021}]%
        {introduceVVC}
\bibfield{author}{\bibinfo{person}{Frank {Bossen}}, \bibinfo{person}{Xiang
  {Li}}, \bibinfo{person}{Karsten {Suehring}}, \bibinfo{person}{Vadim
  {Seregin}}, {and} \bibinfo{person}{A. {Tourapis}}.}
  \bibinfo{year}{2021}\natexlab{}.
\newblock \showarticletitle{JVET AHG report: test model software development
  (AHG3)}.
\newblock \bibinfo{journal}{\emph{JVET document, JVET-W0003-v1}}
  (\bibinfo{year}{2021}).
\newblock


\bibitem[\protect\citeauthoryear{Gao, Kwong, and Jia}{Gao
  et~al\mbox{.}}{2017}]%
        {GTGAO2017}
\bibfield{author}{\bibinfo{person}{Wei Gao}, \bibinfo{person}{Sam Kwong}, {and}
  \bibinfo{person}{Yuheng Jia}.} \bibinfo{year}{2017}\natexlab{}.
\newblock \showarticletitle{Joint Machine Learning and Game Theory for Rate
  Control in High Efficiency Video Coding}.
\newblock \bibinfo{journal}{\emph{IEEE Transactions on Image Processing}}
  \bibinfo{volume}{26}, \bibinfo{number}{12} (\bibinfo{year}{2017}),
  \bibinfo{pages}{6074--6089}.
\newblock
\urldef\tempurl%
\url{https://doi.org/10.1109/TIP.2017.2745099}
\showDOI{\tempurl}


\bibitem[\protect\citeauthoryear{Guo, Zhu, Xu, and Li}{Guo
  et~al\mbox{.}}{2020}]%
        {R-lambda-guotbc20}
\bibfield{author}{\bibinfo{person}{Hongwei Guo}, \bibinfo{person}{Ce Zhu},
  \bibinfo{person}{Mai Xu}, {and} \bibinfo{person}{Shuai Li}.}
  \bibinfo{year}{2020}\natexlab{}.
\newblock \showarticletitle{Inter-Block Dependency-Based CTU Level Rate Control
  for HEVC}.
\newblock \bibinfo{journal}{\emph{IEEE Transactions on Broadcasting}}
  \bibinfo{volume}{66}, \bibinfo{number}{1} (\bibinfo{year}{2020}),
  \bibinfo{pages}{113--126}.
\newblock
\urldef\tempurl%
\url{https://doi.org/10.1109/TBC.2019.2917402}
\showDOI{\tempurl}


\bibitem[\protect\citeauthoryear{Ho, Jin, Liang, Peng, and Li}{Ho
  et~al\mbox{.}}{2021}]%
        {Videomachinelearning-ho21}
\bibfield{author}{\bibinfo{person}{Yung-Han Ho}, \bibinfo{person}{Guo-Lun Jin},
  \bibinfo{person}{Yun Liang}, \bibinfo{person}{Wen-Hsiao Peng}, {and}
  \bibinfo{person}{Xiaobo Li}.} \bibinfo{year}{2021}\natexlab{}.
\newblock \showarticletitle{A Dual-Critic Reinforcement Learning Framework for
  Frame-Level Bit Allocation in HEVC/H.265}. In \bibinfo{booktitle}{\emph{2021
  Data Compression Conference (DCC)}}. \bibinfo{pages}{13--22}.
\newblock
\urldef\tempurl%
\url{https://doi.org/10.1109/DCC50243.2021.00009}
\showDOI{\tempurl}


\bibitem[\protect\citeauthoryear{Hyun, Lee, and Kim}{Hyun
  et~al\mbox{.}}{2020}]%
        {framelevelHyun}
\bibfield{author}{\bibinfo{person}{Myung~Han Hyun}, \bibinfo{person}{Bumshik
  Lee}, {and} \bibinfo{person}{Munchurl Kim}.} \bibinfo{year}{2020}\natexlab{}.
\newblock \showarticletitle{A Frame-Level Constant Bit-Rate Control Using
  Recursive Bayesian Estimation for Versatile Video Coding}.
\newblock \bibinfo{journal}{\emph{IEEE Access}}  \bibinfo{volume}{8}
  (\bibinfo{year}{2020}), \bibinfo{pages}{227255--227269}.
\newblock
\urldef\tempurl%
\url{https://doi.org/10.1109/ACCESS.2020.3046043}
\showDOI{\tempurl}


\bibitem[\protect\citeauthoryear{Li, Li, Li, and Zhang}{Li
  et~al\mbox{.}}{2014}]%
        {slidewindowsLi2014}
\bibfield{author}{\bibinfo{person}{Bin Li}, \bibinfo{person}{Houqiang Li},
  \bibinfo{person}{Li Li}, {and} \bibinfo{person}{Jinlei Zhang}.}
  \bibinfo{year}{2014}\natexlab{}.
\newblock \showarticletitle{$\lambda $ Domain Rate Control Algorithm for High
  Efficiency Video Coding}.
\newblock \bibinfo{journal}{\emph{IEEE Transactions on Image Processing}}
  \bibinfo{volume}{23}, \bibinfo{number}{9} (\bibinfo{year}{2014}),
  \bibinfo{pages}{3841--3854}.
\newblock
\urldef\tempurl%
\url{https://doi.org/10.1109/TIP.2014.2336550}
\showDOI{\tempurl}


\bibitem[\protect\citeauthoryear{{Li}, {Li}, {Li}, and {Zhang}}{{Li}
  et~al\mbox{.}}{2020}]%
        {R-D-2012Li}
\bibfield{author}{\bibinfo{person}{Bin {Li}}, \bibinfo{person}{Houqiang {Li}},
  \bibinfo{person}{Li {Li}}, {and} \bibinfo{person}{Jinlei {Zhang}}.}
  \bibinfo{year}{2020}\natexlab{}.
\newblock \showarticletitle{VTM common test conditions and software reference
  configurations for SDR video}.
\newblock \bibinfo{journal}{\emph{JCT-VC document, JCTVC-K0103}}
  (\bibinfo{year}{2020}).
\newblock


\bibitem[\protect\citeauthoryear{Li, Li, Li, and Chen}{Li
  et~al\mbox{.}}{2018}]%
        {VideoRD-Li18}
\bibfield{author}{\bibinfo{person}{Li Li}, \bibinfo{person}{Bin Li},
  \bibinfo{person}{Houqiang Li}, {and} \bibinfo{person}{Chang~Wen Chen}.}
  \bibinfo{year}{2018}\natexlab{}.
\newblock \showarticletitle{$\lambda$-Domain Optimal Bit Allocation Algorithm
  for High Efficiency Video Coding}.
\newblock \bibinfo{journal}{\emph{IEEE Transactions on Circuits and Systems for
  Video Technology}} \bibinfo{volume}{28}, \bibinfo{number}{1}
  (\bibinfo{year}{2018}), \bibinfo{pages}{130--142}.
\newblock
\urldef\tempurl%
\url{https://doi.org/10.1109/TCSVT.2016.2598672}
\showDOI{\tempurl}


\bibitem[\protect\citeauthoryear{Li, Li, Liu, and Li}{Li et~al\mbox{.}}{2016}]%
        {VideoRD-Li16}
\bibfield{author}{\bibinfo{person}{Li Li}, \bibinfo{person}{Bin Li},
  \bibinfo{person}{Dong Liu}, {and} \bibinfo{person}{Houqiang Li}.}
  \bibinfo{year}{2016}\natexlab{}.
\newblock \showarticletitle{$\lambda$-Domain Rate Control Algorithm for HEVC
  Scalable Extension}.
\newblock \bibinfo{journal}{\emph{IEEE Transactions on Multimedia}}
  \bibinfo{volume}{18}, \bibinfo{number}{10} (\bibinfo{year}{2016}),
  \bibinfo{pages}{2023--2039}.
\newblock
\urldef\tempurl%
\url{https://doi.org/10.1109/TMM.2016.2595264}
\showDOI{\tempurl}


\bibitem[\protect\citeauthoryear{Li, Yan, Li, Liu, and Li}{Li
  et~al\mbox{.}}{2020b}]%
        {R-LAMBDA-Li2020}
\bibfield{author}{\bibinfo{person}{Li Li}, \bibinfo{person}{Ning Yan},
  \bibinfo{person}{Zhu Li}, \bibinfo{person}{Shan Liu}, {and}
  \bibinfo{person}{Houqiang Li}.} \bibinfo{year}{2020}\natexlab{b}.
\newblock \showarticletitle{$\lambda$-Domain Perceptual Rate Control for
  360-Degree Video Compression}.
\newblock \bibinfo{journal}{\emph{IEEE Journal of Selected Topics in Signal
  Processing}} \bibinfo{volume}{14}, \bibinfo{number}{1}
  (\bibinfo{year}{2020}), \bibinfo{pages}{130--145}.
\newblock
\urldef\tempurl%
\url{https://doi.org/10.1109/JSTSP.2019.2963154}
\showDOI{\tempurl}


\bibitem[\protect\citeauthoryear{Li, Liu, Chen, and Liu}{Li
  et~al\mbox{.}}{2020a}]%
        {RC-Li20}
\bibfield{author}{\bibinfo{person}{Yiming Li}, \bibinfo{person}{Zizheng Liu},
  \bibinfo{person}{Zhenzhong Chen}, {and} \bibinfo{person}{Shan Liu}.}
  \bibinfo{year}{2020}\natexlab{a}.
\newblock \showarticletitle{Rate Control For Versatile Video Coding}. In
  \bibinfo{booktitle}{\emph{2020 IEEE International Conference on Image
  Processing (ICIP)}}. \bibinfo{pages}{1176--1180}.
\newblock
\urldef\tempurl%
\url{https://doi.org/10.1109/ICIP40778.2020.9191125}
\showDOI{\tempurl}


\bibitem[\protect\citeauthoryear{Li and Mou}{Li and Mou}{2021}]%
        {TBC21-Li}
\bibfield{author}{\bibinfo{person}{Yang Li} {and} \bibinfo{person}{Xuanqin
  Mou}.} \bibinfo{year}{2021}\natexlab{}.
\newblock \showarticletitle{Joint Optimization for SSIM-Based CTU-Level Bit
  Allocation and Rate Distortion Optimization}.
\newblock \bibinfo{journal}{\emph{IEEE Transactions on Broadcasting}}
  \bibinfo{volume}{67}, \bibinfo{number}{2} (\bibinfo{year}{2021}),
  \bibinfo{pages}{500--511}.
\newblock
\urldef\tempurl%
\url{https://doi.org/10.1109/TBC.2021.3068871}
\showDOI{\tempurl}


\bibitem[\protect\citeauthoryear{Liu and Chen}{Liu and Chen}{2021}]%
        {TCSVT21-Liu}
\bibfield{author}{\bibinfo{person}{Feiyang Liu} {and}
  \bibinfo{person}{Zhenzhong Chen}.} \bibinfo{year}{2021}\natexlab{}.
\newblock \showarticletitle{Multi-Objective Optimization of Quality in VVC Rate
  Control for Low-Delay Video Coding}.
\newblock \bibinfo{journal}{\emph{IEEE Transactions on Image Processing}}
  \bibinfo{volume}{30} (\bibinfo{year}{2021}), \bibinfo{pages}{4706--4718}.
\newblock
\urldef\tempurl%
\url{https://doi.org/10.1109/TIP.2021.3072225}
\showDOI{\tempurl}


\bibitem[\protect\citeauthoryear{Mao, Wang, Wang, and Kwong}{Mao
  et~al\mbox{.}}{2021}]%
        {Rc-mao21}
\bibfield{author}{\bibinfo{person}{Yunhao Mao}, \bibinfo{person}{Meng Wang},
  \bibinfo{person}{Shiqi Wang}, {and} \bibinfo{person}{Sam Kwong}.}
  \bibinfo{year}{2021}\natexlab{}.
\newblock \showarticletitle{High Efficiency Rate Control for Versatile Video
  Coding Based on Composite Cauchy Distribution}.
\newblock \bibinfo{journal}{\emph{IEEE Transactions on Circuits and Systems for
  Video Technology}} (\bibinfo{year}{2021}), \bibinfo{pages}{1--1}.
\newblock
\urldef\tempurl%
\url{https://doi.org/10.1109/TCSVT.2021.3093315}
\showDOI{\tempurl}


\bibitem[\protect\citeauthoryear{Nash~Jr}{Nash~Jr}{1950}]%
        {GTNash50}
\bibfield{author}{\bibinfo{person}{John~F Nash~Jr}.}
  \bibinfo{year}{1950}\natexlab{}.
\newblock \showarticletitle{The bargaining problem}.
\newblock \bibinfo{journal}{\emph{Econometrica: Journal of the econometric
  society}} (\bibinfo{year}{1950}), \bibinfo{pages}{155--162}.
\newblock


\bibitem[\protect\citeauthoryear{Suehring}{Suehring}{2021}]%
        {VVC}
\bibfield{author}{\bibinfo{person}{Karsten Suehring}.}
  \bibinfo{year}{2021}\natexlab{}.
\newblock \bibinfo{booktitle}{\emph{VVC software VTM-13.0}}.
\newblock
\urldef\tempurl%
\url{https://vcgit.hhi.fraunhofer.de/jvet/VVCSoftware_VTM/-/tags/VTM-13.0}
\showURL{%
\tempurl}


\bibitem[\protect\citeauthoryear{Sullivan, Ohm, Han, and Wiegand}{Sullivan
  et~al\mbox{.}}{2012}]%
        {introduceH265}
\bibfield{author}{\bibinfo{person}{Gary~J. Sullivan},
  \bibinfo{person}{Jens-Rainer Ohm}, \bibinfo{person}{Woo-Jin Han}, {and}
  \bibinfo{person}{Thomas Wiegand}.} \bibinfo{year}{2012}\natexlab{}.
\newblock \showarticletitle{Overview of the High Efficiency Video Coding (HEVC)
  Standard}.
\newblock \bibinfo{journal}{\emph{IEEE Transactions on Circuits and Systems for
  Video Technology}} \bibinfo{volume}{22}, \bibinfo{number}{12}
  (\bibinfo{year}{2012}), \bibinfo{pages}{1649--1668}.
\newblock
\urldef\tempurl%
\url{https://doi.org/10.1109/TCSVT.2012.2221191}
\showDOI{\tempurl}


\bibitem[\protect\citeauthoryear{Wang, Ngan, and Li}{Wang
  et~al\mbox{.}}{2016}]%
        {tip-wang16}
\bibfield{author}{\bibinfo{person}{Miaohui Wang}, \bibinfo{person}{King~Ngi
  Ngan}, {and} \bibinfo{person}{Hongliang Li}.}
  \bibinfo{year}{2016}\natexlab{}.
\newblock \showarticletitle{Low-Delay Rate Control for Consistent Quality Using
  Distortion-Based Lagrange Multiplier}.
\newblock \bibinfo{journal}{\emph{IEEE Transactions on Image Processing}}
  \bibinfo{volume}{25}, \bibinfo{number}{7} (\bibinfo{year}{2016}),
  \bibinfo{pages}{2943--2955}.
\newblock
\urldef\tempurl%
\url{https://doi.org/10.1109/TIP.2016.2552646}
\showDOI{\tempurl}


\bibitem[\protect\citeauthoryear{Wang, Wang, Li, Zhang, Wang, and Ma}{Wang
  et~al\mbox{.}}{2020}]%
        {RC-Wang20}
\bibfield{author}{\bibinfo{person}{Meng Wang}, \bibinfo{person}{Shiqi Wang},
  \bibinfo{person}{Junru Li}, \bibinfo{person}{Li Zhang}, \bibinfo{person}{Yue
  Wang}, {and} \bibinfo{person}{Siwei Ma}.} \bibinfo{year}{2020}\natexlab{}.
\newblock \showarticletitle{SSIM Motivated Quality Control for Versatile Video
  Coding}. In \bibinfo{booktitle}{\emph{2020 Asia-Pacific Signal and
  Information Processing Association Annual Summit and Conference (APSIPA
  ASC)}}. \bibinfo{pages}{1122--1127}.
\newblock


\bibitem[\protect\citeauthoryear{Wang, Kwong, Xu, and Zhang}{Wang
  et~al\mbox{.}}{2014}]%
        {gt-wangtip14}
\bibfield{author}{\bibinfo{person}{Xu Wang}, \bibinfo{person}{Sam Kwong},
  \bibinfo{person}{Long Xu}, {and} \bibinfo{person}{Yun Zhang}.}
  \bibinfo{year}{2014}\natexlab{}.
\newblock \showarticletitle{Generalized Nash Bargaining Solution to Rate
  Control Optimization for Spatial Scalable Video Coding}.
\newblock \bibinfo{journal}{\emph{IEEE Transactions on Image Processing}}
  \bibinfo{volume}{23}, \bibinfo{number}{9} (\bibinfo{year}{2014}),
  \bibinfo{pages}{4010--4021}.
\newblock
\urldef\tempurl%
\url{https://doi.org/10.1109/TIP.2014.2341951}
\showDOI{\tempurl}


\bibitem[\protect\citeauthoryear{Wang, Kwong, and Zhang}{Wang
  et~al\mbox{.}}{2013}]%
        {GTwang2013}
\bibfield{author}{\bibinfo{person}{Xu Wang}, \bibinfo{person}{Sam Kwong}, {and}
  \bibinfo{person}{Yun Zhang}.} \bibinfo{year}{2013}\natexlab{}.
\newblock \showarticletitle{Applying Game Theory to Rate Control Optimization
  for Hierarchical B-Pictures}.
\newblock \bibinfo{journal}{\emph{IEEE Transactions on Broadcasting}}
  \bibinfo{volume}{59}, \bibinfo{number}{4} (\bibinfo{year}{2013}),
  \bibinfo{pages}{591--601}.
\newblock
\urldef\tempurl%
\url{https://doi.org/10.1109/TBC.2013.2249351}
\showDOI{\tempurl}


\bibitem[\protect\citeauthoryear{Wiegand, Sullivan, Bjontegaard, and
  Luthra}{Wiegand et~al\mbox{.}}{2003}]%
        {introduceH264}
\bibfield{author}{\bibinfo{person}{T. Wiegand}, \bibinfo{person}{G.J.
  Sullivan}, \bibinfo{person}{G. Bjontegaard}, {and} \bibinfo{person}{A.
  Luthra}.} \bibinfo{year}{2003}\natexlab{}.
\newblock \showarticletitle{Overview of the H.264/AVC video coding standard}.
\newblock \bibinfo{journal}{\emph{IEEE Transactions on Circuits and Systems for
  Video Technology}} \bibinfo{volume}{13}, \bibinfo{number}{7}
  (\bibinfo{year}{2003}), \bibinfo{pages}{560--576}.
\newblock
\urldef\tempurl%
\url{https://doi.org/10.1109/TCSVT.2003.815165}
\showDOI{\tempurl}


\bibitem[\protect\citeauthoryear{Zhao, Lin, Song, Wang, and Niu}{Zhao
  et~al\mbox{.}}{2021}]%
        {GTACMMM-Zhao21}
\bibfield{author}{\bibinfo{person}{Tiesong Zhao}, \bibinfo{person}{Jielian
  Lin}, \bibinfo{person}{Yanjie Song}, \bibinfo{person}{Xu Wang}, {and}
  \bibinfo{person}{Yuzhen Niu}.} \bibinfo{year}{2021}\natexlab{}.
\newblock \showarticletitle{Game Theory-driven Rate Control for 360-Degree
  Video Coding}. In \bibinfo{booktitle}{\emph{Proceedings of the 29th ACM
  International Conference on Multimedia}}. \bibinfo{pages}{3998--4006}.
\newblock


\bibitem[\protect\citeauthoryear{Zhao, Wang, Wang, and Chen}{Zhao
  et~al\mbox{.}}{2015}]%
        {VideoSSIM-Zhao15}
\bibfield{author}{\bibinfo{person}{Tiesong Zhao}, \bibinfo{person}{Jiheng
  Wang}, \bibinfo{person}{Zhou Wang}, {and} \bibinfo{person}{Chang~Wen Chen}.}
  \bibinfo{year}{2015}\natexlab{}.
\newblock \showarticletitle{SSIM-Based Coarse-Grain Scalable Video Coding}.
\newblock \bibinfo{journal}{\emph{IEEE Transactions on Broadcasting}}
  \bibinfo{volume}{61}, \bibinfo{number}{2} (\bibinfo{year}{2015}),
  \bibinfo{pages}{210--221}.
\newblock
\urldef\tempurl%
\url{https://doi.org/10.1109/TBC.2015.2424012}
\showDOI{\tempurl}


\bibitem[\protect\citeauthoryear{Zhou, Wei, Kwong, Jia, and Fang}{Zhou
  et~al\mbox{.}}{2020a}]%
        {VideoJND-Zhou20}
\bibfield{author}{\bibinfo{person}{Mingliang Zhou}, \bibinfo{person}{Xuekai
  Wei}, \bibinfo{person}{Sam Kwong}, \bibinfo{person}{Weijia Jia}, {and}
  \bibinfo{person}{Bin Fang}.} \bibinfo{year}{2020}\natexlab{a}.
\newblock \showarticletitle{Just Noticeable Distortion-Based Perceptual Rate
  Control in HEVC}.
\newblock \bibinfo{journal}{\emph{IEEE Transactions on Image Processing}}
  \bibinfo{volume}{29} (\bibinfo{year}{2020}), \bibinfo{pages}{7603--7614}.
\newblock
\urldef\tempurl%
\url{https://doi.org/10.1109/TIP.2020.3004714}
\showDOI{\tempurl}


\bibitem[\protect\citeauthoryear{Zhou, Wei, Kwong, Jia, and Fang}{Zhou
  et~al\mbox{.}}{2021}]%
        {Videomachinelearning-Zhou21}
\bibfield{author}{\bibinfo{person}{Mingliang Zhou}, \bibinfo{person}{Xuekai
  Wei}, \bibinfo{person}{Sam Kwong}, \bibinfo{person}{Weijia Jia}, {and}
  \bibinfo{person}{Bin Fang}.} \bibinfo{year}{2021}\natexlab{}.
\newblock \showarticletitle{Rate Control Method Based on Deep Reinforcement
  Learning for Dynamic Video Sequences in HEVC}.
\newblock \bibinfo{journal}{\emph{IEEE Transactions on Multimedia}}
  \bibinfo{volume}{23} (\bibinfo{year}{2021}), \bibinfo{pages}{1106--1121}.
\newblock
\urldef\tempurl%
\url{https://doi.org/10.1109/TMM.2020.2992968}
\showDOI{\tempurl}


\bibitem[\protect\citeauthoryear{Zhou, Wei, Wang, Kwong, Fong, Wong, and
  Yuen}{Zhou et~al\mbox{.}}{2020b}]%
        {R-LAMBDA-ZHOU2020}
\bibfield{author}{\bibinfo{person}{Mingliang Zhou}, \bibinfo{person}{Xuekai
  Wei}, \bibinfo{person}{Shiqi Wang}, \bibinfo{person}{Sam Kwong},
  \bibinfo{person}{Chi-Keung Fong}, \bibinfo{person}{Peter H.~W. Wong}, {and}
  \bibinfo{person}{Wilson Y.~F. Yuen}.} \bibinfo{year}{2020}\natexlab{b}.
\newblock \showarticletitle{Global Rate-Distortion Optimization-Based Rate
  Control for HEVC HDR Coding}.
\newblock \bibinfo{journal}{\emph{IEEE Transactions on Circuits and Systems for
  Video Technology}} \bibinfo{volume}{30}, \bibinfo{number}{12}
  (\bibinfo{year}{2020}), \bibinfo{pages}{4648--4662}.
\newblock
\urldef\tempurl%
\url{https://doi.org/10.1109/TCSVT.2019.2959807}
\showDOI{\tempurl}


\end{thebibliography}

\end{document}